\documentclass[sigconf]{acmart}

\usepackage{booktabs} 
\usepackage{graphicx}
\usepackage{subfig}
\usepackage{multirow}

\setcopyright{acmcopyright}

\newcommand\ainote[1]{\textcolor{red}{[AI: #1]}}

\newcommand\shnote[1]{\textcolor{blue}{SH: #1}}

\newcommand{\ignore}[1]{}

\begin{document}
\title{Cascade-LSTM: Predicting Information Cascades using\\ Deep Neural Networks}

\author{Sameera Horawalavithana}
\orcid{0000-0002-0327-3819}
\affiliation{%
  \institution{Department of Computer Science and Eng., University of South Florida}
  \streetaddress{E. Fowler Ave.}
  \city{Tampa}
  \state{FL}
  \postcode{33620}
  \country{USA}}
\email{sameera1@mail.usf.edu}





\author{John Skvoretz}
\affiliation{%
  \institution{Department of Sociology, University of South Florida}
  \streetaddress{E. Fowler Ave.}
  \city{Tampa}
  \state{FL}
  \postcode{33620}
  \country{USA}}
\email{jskvoretz@usf.edu}


\author{Adriana Iamnitchi}
\affiliation{%
  \institution{Department of Computer Science and Eng., University of South Florida}
  \streetaddress{E. Fowler Ave.}
  \city{Tampa}
  \state{FL}
  \postcode{33620}
  \country{USA}}
\email{anda@cse.usf.edu}

\renewcommand{\shortauthors}{S. Horawalavithana et al.}

\begin{abstract}
Predicting the flow of information in dynamic social environments is relevant to many areas of the contemporary society, from disseminating health care messages to meme tracking.
While predicting the growth of information cascades has been successfully addressed in diverse social platforms, predicting the temporal and topological structure of information cascades has seen limited exploration. 
However, accurately predicting how many users will transmit the message of a particular user and at what time is paramount for designing practical intervention techniques. 

This paper leverages Long-Short Term Memory (LSTM) neural network techniques to predict two spatio-temporal properties of information cascades, namely the \textit{size} and \textit{speed} of individual-level information transmissions.
We combine these prediction algorithms with probabilistic generation of cascade trees into a generative test model that is able to accurately generate cascade trees in two different platforms, Reddit and Github. 
Our approach leads to a classification accuracy of over 73\% for information transmitters and 83\% for early transmitters in a variety of social platforms.  

\end{abstract}

%
%



\maketitle

\section{Introduction}
\label{sec:intro}

Understanding how information is disseminated in online social environments has significant real world impact, from health care to marketing. 
Significant effort has been invested in predicting different properties of information cascades, such as size~\cite{yu2015micro}, temporal growth~\cite{li2017deepcas}, and virality~\cite{cheng2014can}. 
At the same time, effort has been invested in characterizing information cascades in various platforms.
For example, Cheng et al.~\cite{cheng2018diffusion} characterized the types of information cascades in Facebook. 
Blackburn et al.~\cite{Blackburn2014CheatingOnline} studied the social contagion of cheating behavior in online gaming platforms.  

One significant question is how to intervene in information cascades. 
Vosoughi et al.~\cite{vosoughi2018spread} determined based on a collection of tweets of political news that false information spreads faster, farther, deeper and broader than true facts, even when controlling for network structure characteristics such as number of followers or for the existence of bots in the platform. 
This phenomenon may be explained by human factors such as emotional reaction to surprise, fear and disgust that are more likely induced by fabricated news.  
Given that 60\% of the population in the US is taking their news from social media, such phenomena can be disturbing to the functioning of the society.

One approach to mitigate these risks is to develop intervention techniques to contain the spread of bad content or behavior. 
News fact-checking is one such technique meant at intervening in the spread of misinformation. 
However, empirical observations show that fact-checks do not travel to the same distance and on the same path as the rumors did, basically because users who reacted emotionally to the rumors and spread them will not bother to spread the fact checking information. 
One possibility to address this dissonance is to prevent the further spread of false information by a technique similar to vaccination. 
The question becomes that of predicting the shape over time of the information cascade, to be able to ``innoculate'' users by proving them the fact checking information before they receive the rumor, thus discouraging them from spreading it farther. 
What is missing in the plan are techniques that accurately predict the shape of the information cascades.

This work advances the state of the art in predicting the structure of information cascades over time by employing deep neural network techniques. 
The main contribution of this paper is a generative approach that leverages Long-Short Term Memory (LSTM) techniques to predict the topological structure over time of information cascade trees. 
We show that our methodology is useful for predicting information cascades of different spatio-temporal patterns of growth. 
Specifically, we test our approach on two opposite types of platforms: a platform with slowly growing, shallow cascades (Github), and a platform with fast growing, potentially tall and broad cascades (Reddit). 
In Reddit, a massive collection of forums, users engage with content posted by other users via comments that are typically fast-paced but over relatively short periods. In GitHub, a collaborative software development platform, users collaboratively engage with software repositories via a diversity of actions, such as issuing comments, contributing code fixes, watching the evolution of repositories, or copying (forking) repositories in a much slower pace and over significantly longer periods of time.

In our approach, we represent cascade trees in a data model consists of a rich feature set.
Our empirical evaluation demonstrate that this approach can accurately predict the shape of the cascade tree over time (that is, number of levels, number of nodes per level, virality, and the time associated with every node's adoption).
The proposed model is also compared with several baseline models in both cascade prediction and generation tasks.
We significantly improve the prediction accuracy over the baseline models, and provide plausible explanations for the success.

An overview of cascade prediction tasks in online environments is presented in Section~\ref{sec:related}. 
Section~\ref{sec:methodology} describes our LSTM-based generative test approach. 
Section~\ref{sec:dataset} presents the characteristics of the two platforms we experiment with, Reddit and GitHub, and Section~\ref{sec:results} presents the accuracy of temporal growth predictions on these two platforms. 
The paper concludes with a summary and discussion of our findings. 

\ignore{

\ainote{Maybe the story can be motivated by the following: while understanding how quickly information spreads in online environments is paramount in many areas, from viral marketing to disseminating health information, predicting where information spreads over time -- thus, the "shape" of an information cascade -- is essential for intervention techniques. For example, it has been shown that information spreads much wider than its validation from fact checkers -- is there a story here? }
Predicting the information spread in online environments has seen significant interest in areas such as health management, product marketing, and fact checking.

Outline:

Big important problem for many domains. 

Useful for intervention techniques. In particular, knowing where and when information is spreading is particularly important.

However, cascades are notoriously unpredictable. 

In the last years, however, significant progress has been made in predicting important properties of information cascades  in some domains. For example, size ... Anything else?
\shnote{size, growth and shape}

\shnote{"To understand the underlying structure of an event, therefore, one must reconstruct the full adoption cascade, which in turn requires observing both individual-level adoption decisions and also the social ties over which these adoptions spread."\cite{goel2015structural}}

However, the question of where and when information spreads is still unanswered. 
}

\section{Related Work}
\label{sec:related}


Information cascades are ubiquitous in online social environments.
Such cascades capture how content or behaviors propagate over an underlying network~\cite{easley2010networks}.
In traditional cascading environments, an individual (i.e., adopter) can be influenced by neighbors or by a community that the individual belongs to, thus assuming an underlying social network topological structure that both restricts and facilitates information diffusion.

However, sometimes the underlying network is not visible, even in the digital world where everything seems to be public or retrievable.
For example, Twitter has the follower-followee network, but its broadcasting structure~\cite{Kwak:2010:TSN:1772690.1772751} is too unrestricted to give sufficient information to guide the information diffusion. 
Gomez et al.~\cite{Gomez-Rodriguez:2012:IND:2086737.2086741} recognized this problem (more prevalent in contexts outside the digital world) and proposed heuristics to recreate the underlying network from observed paths of diffusion. 
In many social platforms, however, the flow of information is not restricted to the underlying social network topology.
Predicting the evolution of a conversation thread over time (traditionally seen as a conversation cascade, in which the behavior adopted is of contributing to the conversation thread) in such an environment is thus technically challenging. 

Yet various macroscopic properties of cascades were inferred even under such conditions. 
The size of a cascade in the future was predicted in numerous previous studies.
Several works adopt statistical approaches~\cite{zhao2015seismic,liben2008tracing}, while others used machine-learning methods with domain-specific features~\cite{cheng2014can,yu2015micro,kupavskii2012prediction}.
DeepCas~\cite{li2017deepcas} proposed a framework to predict the incremental growth of a cascade.
They use a network-embedding approach based on independent cascade paths. 
Cheng et. al.~\cite{cheng2014can} distinguish viral cascades in a data-set of Facebook photo re-shares. For this task, both temporal and structural features are important to the prediction task.

Several individual-level prediction tasks are studied over state-of-the-art machine-learning algorithms.
Given a Twitter user, the prediction tasks of performing a retweet~\cite{petrovic2011rt} or sharing a URL~\cite{galuba2010outtweeting} are studied using more sophisticated linear models.
Topo-LSTM~\cite{wang2017topological} is recent work which proposed a new data model to represent cascades in a structure of a directed acyclic graph (DAG).
However, they do not assume a user to have multiple actions in the same cascade.

In our setup, we support such situation by treating cascades as trees with nodes ordered by activation time.
Further, we use a rich set of content features to represent nodes participate in the cascade.
These features enable us to learn an implicit semantic structure of the content being diffused.
We also predict the rate of individual adoption to accurately measure the lifetime of the cascade as a function of size.

Many theoretical models have been proposed to model complex contagion (e.g., Bass~\cite{bass1969new}, SIR~\cite{kermack2003contribution}).
Theoretical models make the assumption of an underlying network that governs the cascade process.
Generative models also assume an underlying network structure, 
In~\cite{Ghosh:2011:FQA:1935826.1935917}, Ghosh and Lerman introduce a mathematical framework to quantify and analyze cascades. 
Our baseline generative models are constructed according to Cheng et. al.~\cite{cheng2018diffusion} who found that the branching factor (conditional on degree and level) can be used to effectively model cascade trees.




\section{Methodology}
\label{sec:methodology}
We introduce a generative mechanism to predict the temporal structure of a cascade with the support of probabilistic models and Long-Short Term Memory (LSTM) neural networks.
Our approach is to train a machine-learning algorithm to act as a filter for identifying realistic cascades for a particular platform from a large pool of generated cascades.
We use generative techniques (inspired from~\cite{cheng2018diffusion} and described in Section~\ref{sec:cascade-generation}) to generate this pool of cascades, and use the LSTM-based filter (described in Section~\ref{sec:cascade-LSTM}) to rank the cascades according to an accuracy metric.
For training the LSTM filter, we use the data representation presented in Section~\ref{sec:cascade-data-model}. 

\subsection{Cascade Generation}
\label{sec:cascade-generation}

We employ a probability-based cascade generation approach~\cite{cheng2018diffusion} to construct the pool of cascades from which to select the best.

Baseline models are based on three conditional probability distributions drawn from the observed cascades in the training period.
First distribution conditions the degree (i.e., number of adoptions), and the second distribution conditions the semantic values of the content (e.g., sentiment score of a Reddit comment) of an individual node by the level of the cascade tree.
Third distribution conditions the sequence of adoption delays by the size of the cascade.
We build the cascade trees recursively where the nodes are drawn from three conditional probability distributions.
We also use cascades generated with this approach to benchmark our results.

In an empirical analysis, we found such techniques fail to accurately predict the adoption/ spread decisions of individuals participate in the cascade.
Specifically, cascades generated with the conditional degree based model~\cite{cheng2018diffusion} tend to be shallow and larger.

\ignore{
First, we outline the data representation for the nodes participate in the cascade.
Second, we describe the proposed machine-learning setup which includes the details of the prediction objectives.
Third, we describe the techniques for the generation of cascade structure over time.
Finally, we present a set of performance metrics to evaluate the goodness of generated cascade structures.
}



		
	 

\subsection{Cascade Data Model}
\label{sec:cascade-data-model}

Several studies in the past model cascades based on an underlying network.
For example, retweet cascades are studied over the follower network of Twitter users~\cite{wang2017topological,li2017deepcas} and resharing cascades in Facebook are constructed over the friendship network~\cite{cheng2014can}.
Such a selection of an underlying network is domain-specific and, intuitively, can contribute significantly to the accuracy of predicting cascades.
However, in some platforms, such underlying networks are not visible.
Our empirical analysis shows that most of the direct user interactions in Reddit occurred only once.
On the other hand, the number of new users appeared in Reddit is proportionately high.

\begin{figure*}[htbp]
	\centering
	\centering
\begin{tabular}{@{}ccc@{}}
	\subfloat[Cascade Tree]{
		\includegraphics[width=0.325\textwidth]{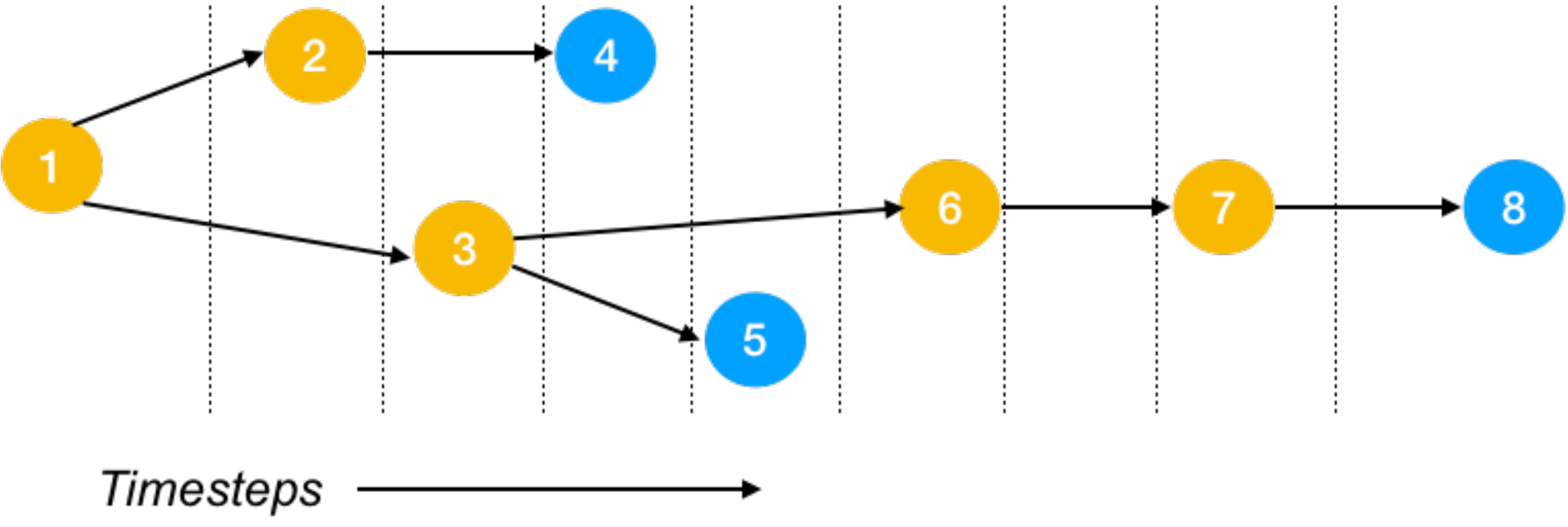}
			\label{fig:cascade_tree}
	}
&
	
	\subfloat[A Brick of Single Cascade]{
		\includegraphics[width=0.325\textwidth]{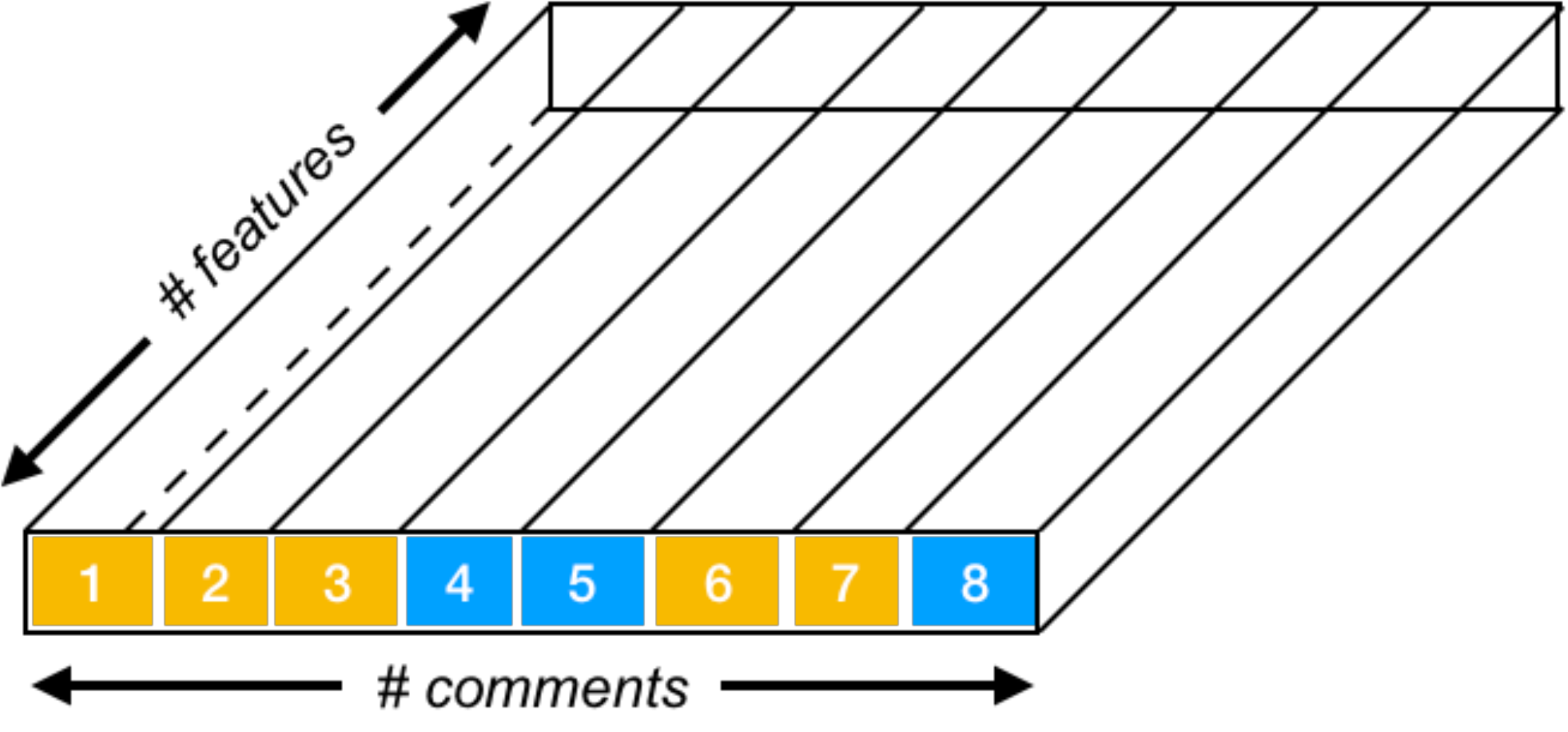}
			\label{fig:cascade_brick}
	}
	
&
	
	\subfloat[A Block of Multiple Cascades]{
		\includegraphics[width=0.325\textwidth]{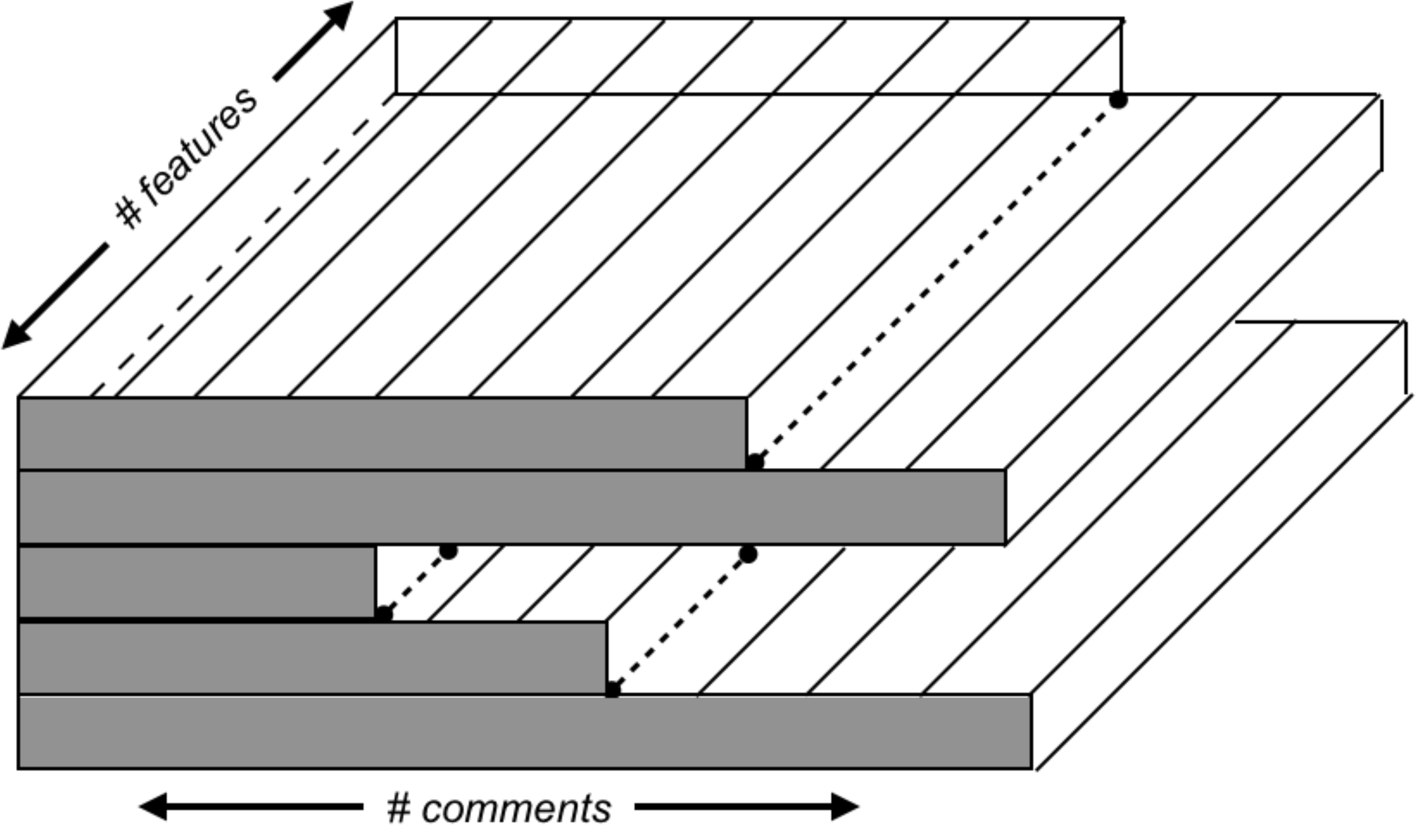}
			\label{fig:cascade_block}
	}
\end{tabular}
	 \caption{Representation of cascade trees in our setup (e.g., an online conversation thread), a) Nodes are ordered by the participation time in a cascade tree, where nodes are highlighted in two colors; yellow and blue for leaf and non-leaf nodes respectively. b) Each node is represented by a spatiotemporal feature vector, and stack together in the order of submission time. c) The block view of multiple cascades, which presents cascades that could have arbitrary number of nodes}
	 \label{fig:cascade_view}
	 
\end{figure*}


Conceptually, we represent a cascade as a tree of messages, as shown in Figure~\ref{fig:cascade_tree}, in which nodes/messages can be a post (for the root node) or comments (for the rest of the nodes in the tree).
Thus, a user can author multiple messages (i.e., nodes) in a particular cascade. 
A link in the cascade tree represents the ``responding to'' relationship. 
Each node/message is described by the time of its posting and a set of other features as described below. 

\ignore{
There are two main reasons to adopt such simplified structure.
\begin{enumerate}
    \item Lack of a clear underlying network to govern the cascade in which no individual has a persistent set of neighbors.
    \item Enable an individual to participate multiple-times in a cascade at different positions.
\end{enumerate}
}


Figure~\ref{fig:cascade_brick} presents the data structure that represents a cascade for our machine learning algorithms. 
Nodes (i.e., messages) are ordered chronologically, by their post time.
Intuitively, each node is described by its author (for example, the Reddit user who posted that message) and the content of the message.
Specifically, each node is described by properties organized in three main categories:i)~spatio-temporal properties, that capture the position of an individual message in a cascade; ii)~user features; and iii)~content features. 
These features are detailed in Table~\ref{tbl:feature_desc}.

We refer to the 2-dimensional feature vector that represents one cascade as a ``brick''. 
We represent multiple cascade bricks stacked together in a block as shown in Figure~\ref{fig:cascade_block} to plug into the machine-learning pipeline.

\subsection{Machine Learning Setup}
\label{sec:cascade-LSTM}

In this section, we describe a set of learning tasks over the representation of cascades described above.
First, we outline the machine-learning model with the details of LSTM setup, including the prediction tasks,
Next we propose a method to generate realistic cascades.

\subsubsection{LSTM-Model}
Traditional machine-learning techniques do not keep the state of predecessors to reason the actions of successors.
Recurrent neural-networks (RNN) are proposed to mitigate such problems, and proved to be useful in sequence learning tasks~\cite{hochreiter1997long}.

For each individual to react in the cascade, the past reactions of predecessors matter.
As an example, the last comment to a Reddit post could trigger an individual to make an immediate reaction.
Meanwhile, such an individual needs to know the details of the overall conversation (i.e., How the conversation unfolded up until his reaction)
Long-Short Term Memory (LSTM) neural networks is a variation of RNN to tackle the context of long-term dependency~\cite{hochreiter1997long}.

As shown in Figure~\ref{fig:cascade_brick}, we feed a cascade brick into the LSTM.
We use the memory-cell design of a standard LSTM in our work~\cite{hochreiter1997long} which is implemented in Keras~\cite{chollet2015keras}.
Our LSTM setup includes two blocks of memory-cells with $32$ and $8$ hidden units, and we use Adam algorithm for the optimization with a learning rate of $0.001$ (based on hyper-parameter optimization).
Cascade bricks are different in shape (e.g., the size of an online conversation), such that we feed bricks one by one to train in LSTM.

\subsubsection{Prediction Tasks}
\label{sec:pred_tasks}
In general, we predict the likelihood of observing a given sequence of adoptions in a cascade.
We use two individual-level properties (e.g., branching factor and speed) of the cascade as the target units for the prediction tasks.
In the first prediction task, we classify the messages as leaves (class 0) or branch (class 1) nodes in the tree.
(Note that these node positions determine the shape of the cascade.) 

The second prediction task classifies messages by the delay with which they are posted in response to their parent.
We refer to this delay as \emph{propagation delay}. 
We consider the median propagation delay within a cascade as the borderline between the two classes: messages with a propagation delay larger than this median are called late adopters (class 1), while the others are early adopters (class 0). 

\subsubsection{Generative Test}
\label{sec:generative_test}
We use the cascade generative approach described in Section~\ref{sec:cascade-generation} to construct a thousand of blocks as represented in Figure~\ref{fig:cascade_block}. 
Specifically, the input to the generator is the original post or repository described by features as detailed in Table~\ref{tbl:feature_desc}, where many of the features describe it as the root of a potential (unknown) cascade. 
Using the generator in Section~\ref{sec:cascade-generation}, a cascade tree is generated probabilistically and represented as a brick like that in Figure~\ref{fig:cascade_brick}.
If multiple cascades "seeds" are given at once, a block (as in Figure~\ref{fig:cascade_block}) is constructed, by stacking all the bricks together. 
The many variants of such blocks are generated in multiple trials. 

Our object now becomes to select the best block according to the learned Cascade-LSTM model. 
In this case, Cascade-LSTM acts as a selector in the following way. 
The trained LSTM model takes as input the chronological sequence of the messages generated as part of the cascade but with all links in the cascade tree removed. 
The trained model will generate the labels that describe whether a message is a branch node or a leaf.
These labels are then compared with those generated probabilistically. 
We calculate accuracy as AUC and rank blocks by the mean accuracy.

\ignore{
\subsection{Performance Metrics}

}

\section{Datasets}
\label{sec:dataset}
For our empirical investigations, we chose real-world datasets from two platforms, Reddit and Github.
Data from both platforms is publicly available. 
We simply used topic-based subsets of these data (to focus our study on issues other than big data challenges) and in some cases augmented data with metrics that capture sentiment, controversiality, etc.

Reddit is a popular discussion forum where users can post textual content, share URLs, and media, etc.
The subset of the Reddit dataset contains thousands of conversation threads that occurred between January 2015 and August 2017 in seven Reddit subreddits. 
Data was extracted from the official crypto-currency subreddits (e.g., /r/bircoin, /r/ethereum, /r/Monero etc.) by filtering the conversation with the keywords appeared in the domain of "crypto-currency".
We extract each thread of conversation and represent it as a cascade tree.
A node in the cascade tree consists of the textual content of a Reddit post/comment and its author.
A pair of nodes (source to target) connected in a directed edge where the direction presents the target node makes a reaction towards the source node. 

We obtained more than 200,000 cascades in Reddit, with a maximum size of 7,868 and a maximum depth of 160 (Table~\ref{tbl:cascade_props}). 
The longest lifespan for a Reddit cascade was identified at 311 days (between the initial post and the time of the last comment in that thread). 

Github is a software development platform where users interact with software repositories via different actions related to the git software versioning system, such as forking, pushing/pulling repository updates, branching, etc.
A direct type of information cascades in Github is copying the existing repositories (via forking), intuitively for the purpose of adapting the original code-base for different purposes or different organizations.
We thus extract fork-based cascade trees, where a particular node in the cascade consists of a software repository and the user (i.e., forkee/ forker).
Directed edge between a pair of nodes (source to target) indicates that the target repository is forked from the source repository.

The majority of Github cascade trees are shallow: the cascade finishes at depth one, where the root repository is forked into other repositories, but none of the resulting repositories gets forked during our 2.5 years observation window.
While this is a realistic behavior in the Github platform, it is not particularly useful for studying cascades: trivially predicting these many and very shallow cascades would lead to meaningless claims of high accuracy. 
In this study we only consider cascade trees of depth greater than one.
We thus identified over 6,000 cascades, with minimum depth 2 and maximum depth 4 and the longest lifespan of 2,940 days (in which the root repository was created in October 2007 with the latest repository forked from it in January 2017). 

The choice for these two datasets is motivated by differences along many axes, as already suggested above.  
Intuitively, however, the following differences capture the contrasting nature of the two platforms: Reddit is a conversation platform, where content, even when of technical nature, can be created relatively fast, and where anybody can participate.
Github, in contrast, being at core a collaborative software versioning system, is targeting a more specialized audience (software developers). 
Content in Github (i.e., software) takes longer to be created/updated, understood, and disseminated.  

Two classes of properties of these datasets are particularly relevant for our empirical study: the properties of the observed cascades in the two platforms, compared in Section~\ref{sec:datasets-cascades}, and the properties of the features described in Table~\ref{tbl:feature_desc}, detailed in Section~\ref{sec:datasets-node-features}.  



\begin{table}[t]
	\centering
	\caption{Properties of cascades in two platforms: Reddit conversations and Github fork trees. For Github we only consider cascades with $depth \ge 2$.
	}
	\label{tbl:cascade_props}
	\begin{tabular}{|l||r|r|}
		\hline 
		Measurement & Reddit & Github \\ \hline \hline
		Number of cascades & 209,721 & 6,072 \\ \hline
		Number of adoptions & 3,580,162 & 39,138 \\ \hline
		Number of unique users & 144,457 & 33,238 \\ \hline
		Max cascade lifetime (days) & 311 & 2,980 \\ \hline
		Max cascade size & 7,868 & 421 \\ \hline 
		Max cascade depth & 160 & 4 \\ \hline 
		Max cascade breadth by level & 7,578 & 417 \\ \hline 
		\end{tabular} 
\end{table}



\subsection{Cascades in Reddit vs. Github}
\label{sec:datasets-cascades}




\begin{figure}[htbp]
	\centering
	\centering
\begin{tabular}{@{}cc@{}}
	\subfloat[Reddit]{
		\includegraphics[width=0.24\textwidth]{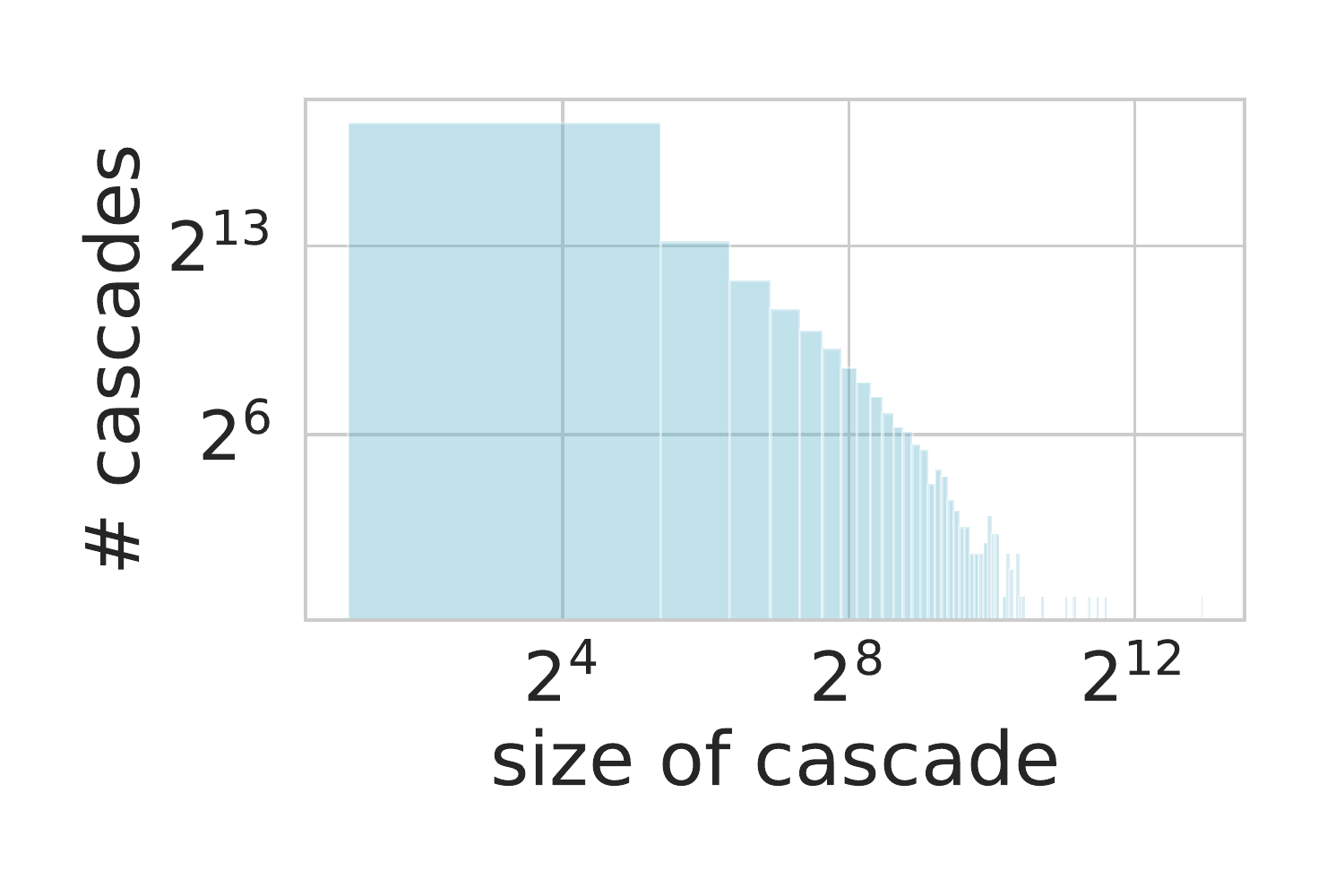}
			\label{fig:crypto_cascade_size}
	}
&
\subfloat[Github]{
		\includegraphics[width=0.24\textwidth]{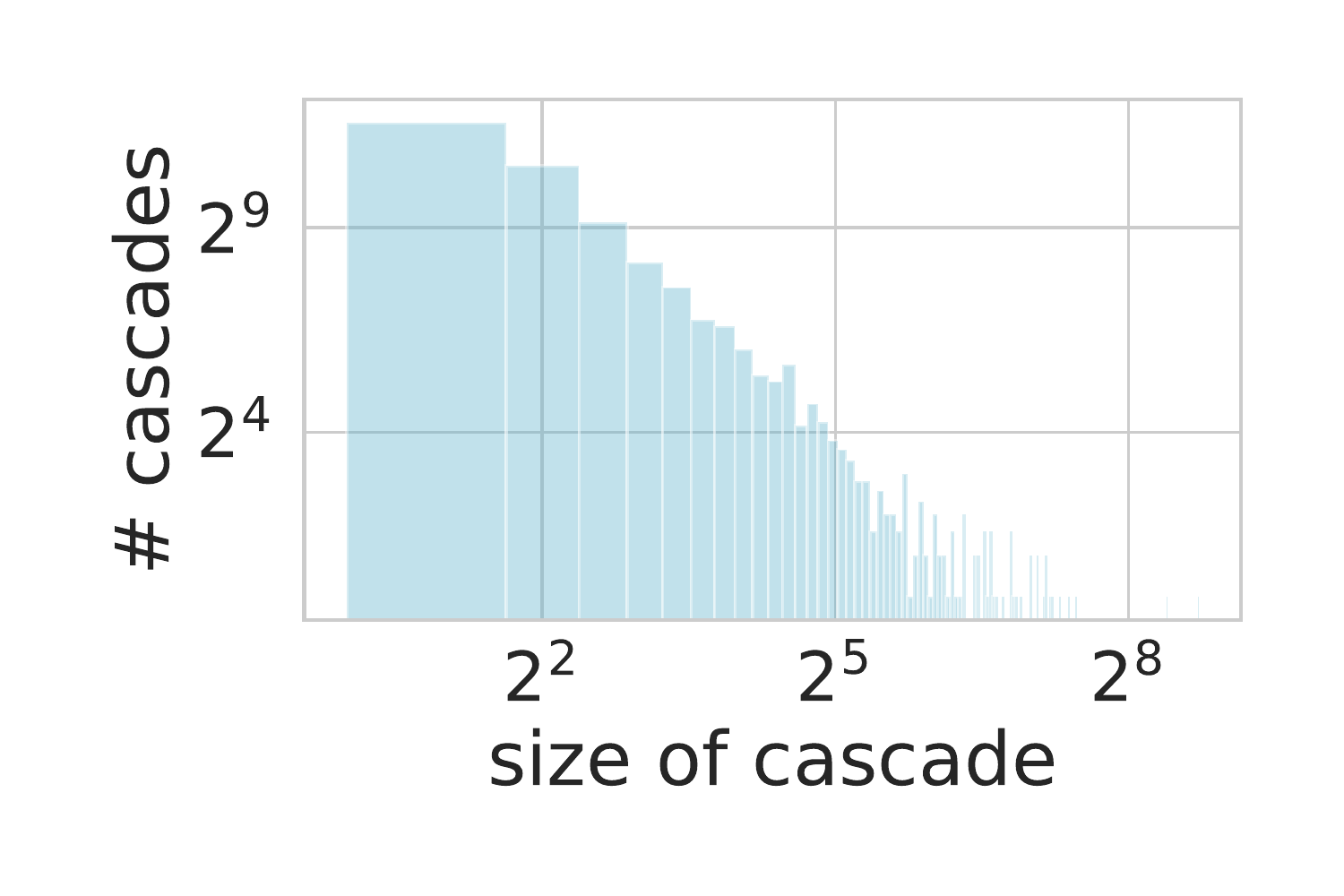}
			\label{fig:github-fork_cascade_size}
	}
	\\
	
		\subfloat[Reddit]{
		\includegraphics[width=0.24\textwidth]{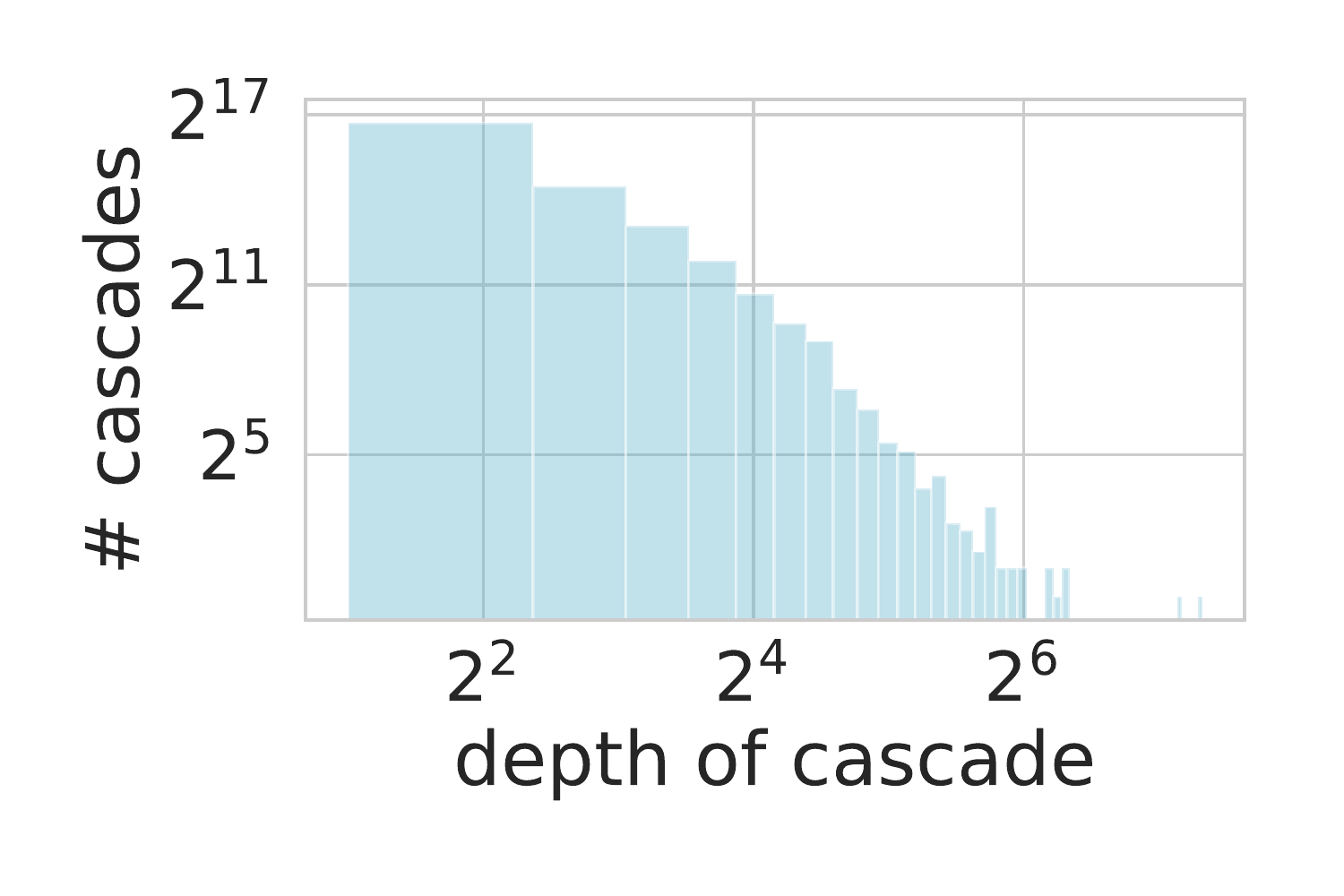}
			\label{fig:crypto_cascade_max_depth}
	}
	
&
	\subfloat[Github]{
		\includegraphics[width=0.24\textwidth]{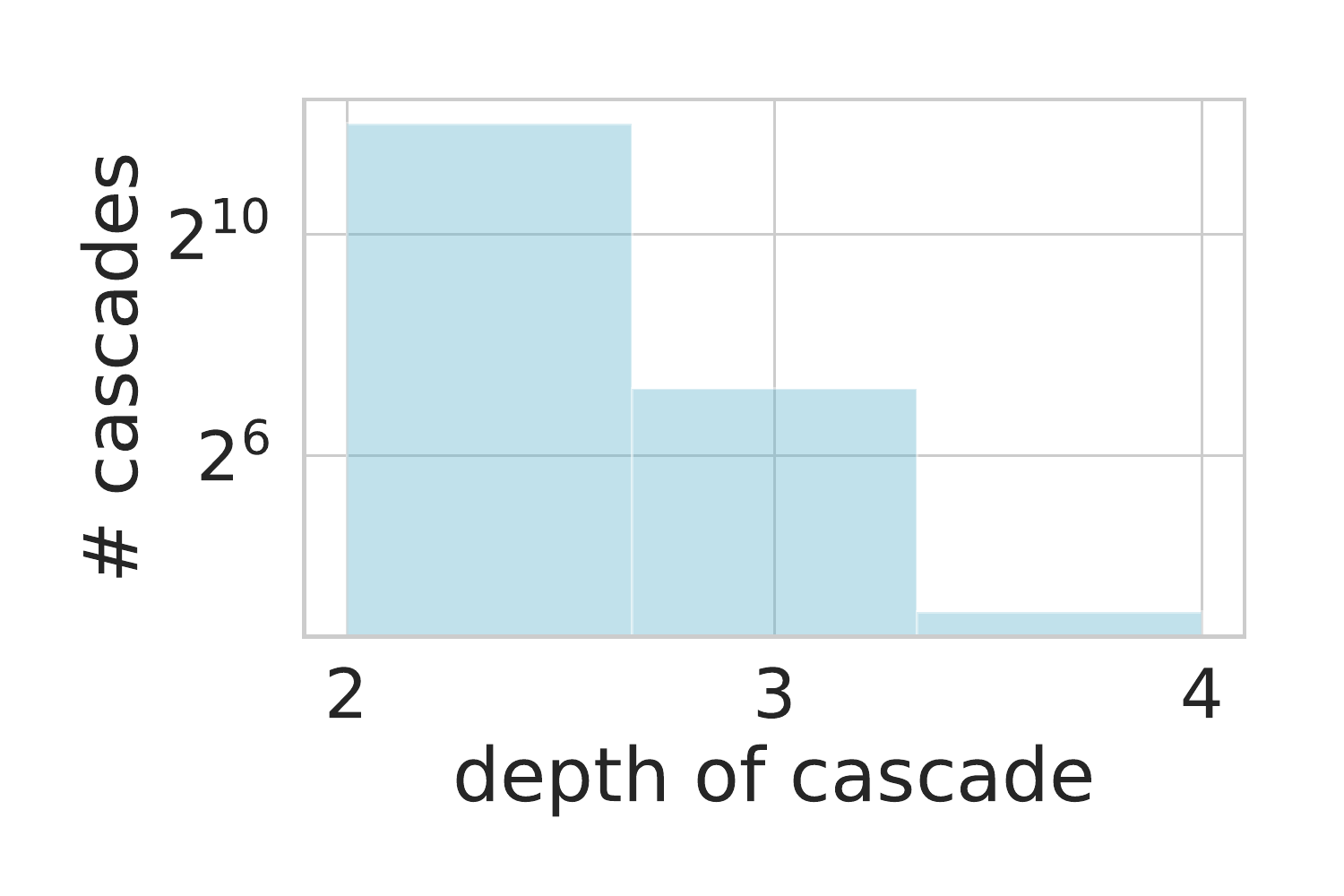}
			\label{fig:github-fork_cascade_max_depth}
	}
	\\
		\subfloat[Reddit]{
		\includegraphics[width=0.24\textwidth]{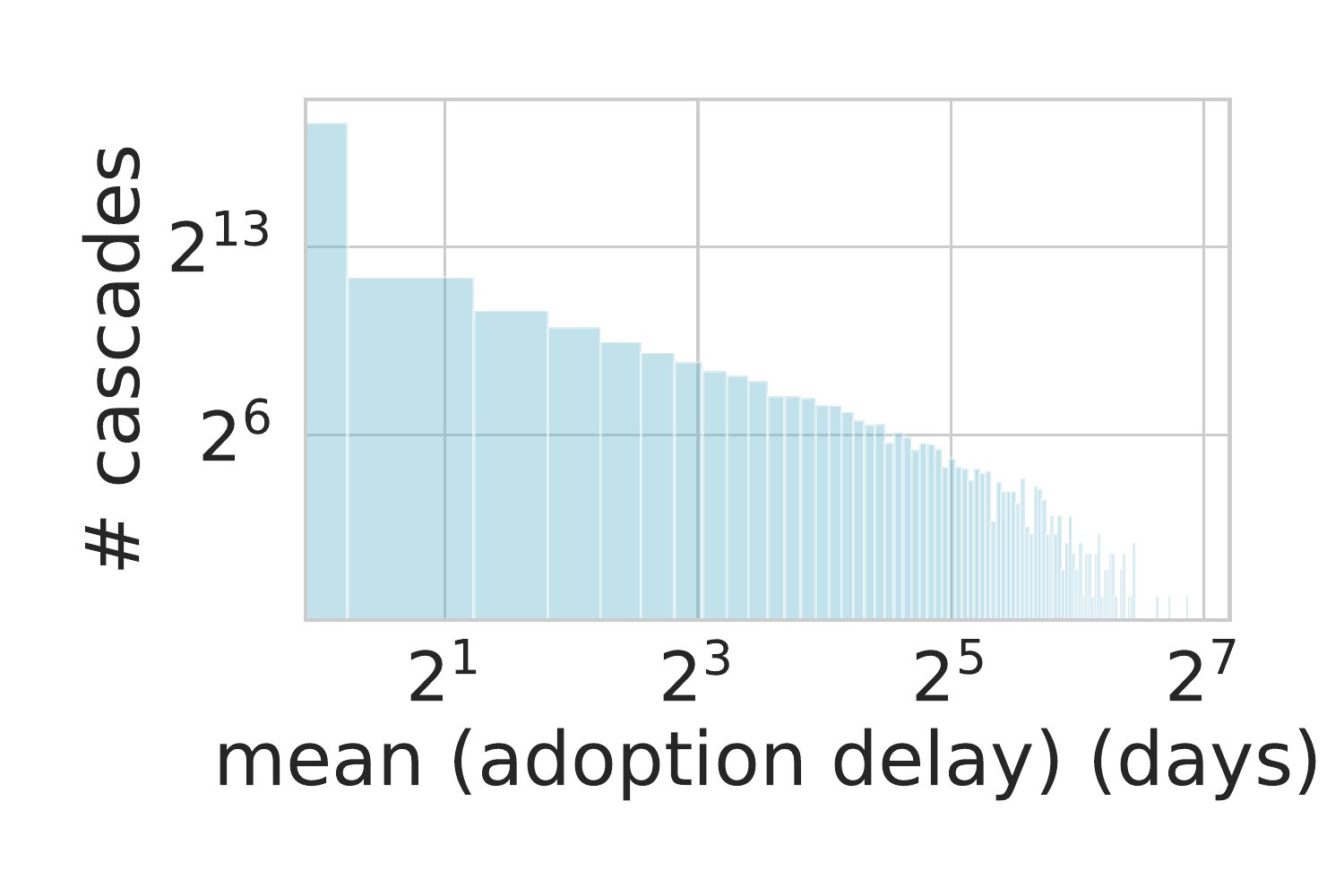}
			\label{fig:crypto_cascade_delay}
	}
	&
	\subfloat[Github]{
		\includegraphics[width=0.24\textwidth]{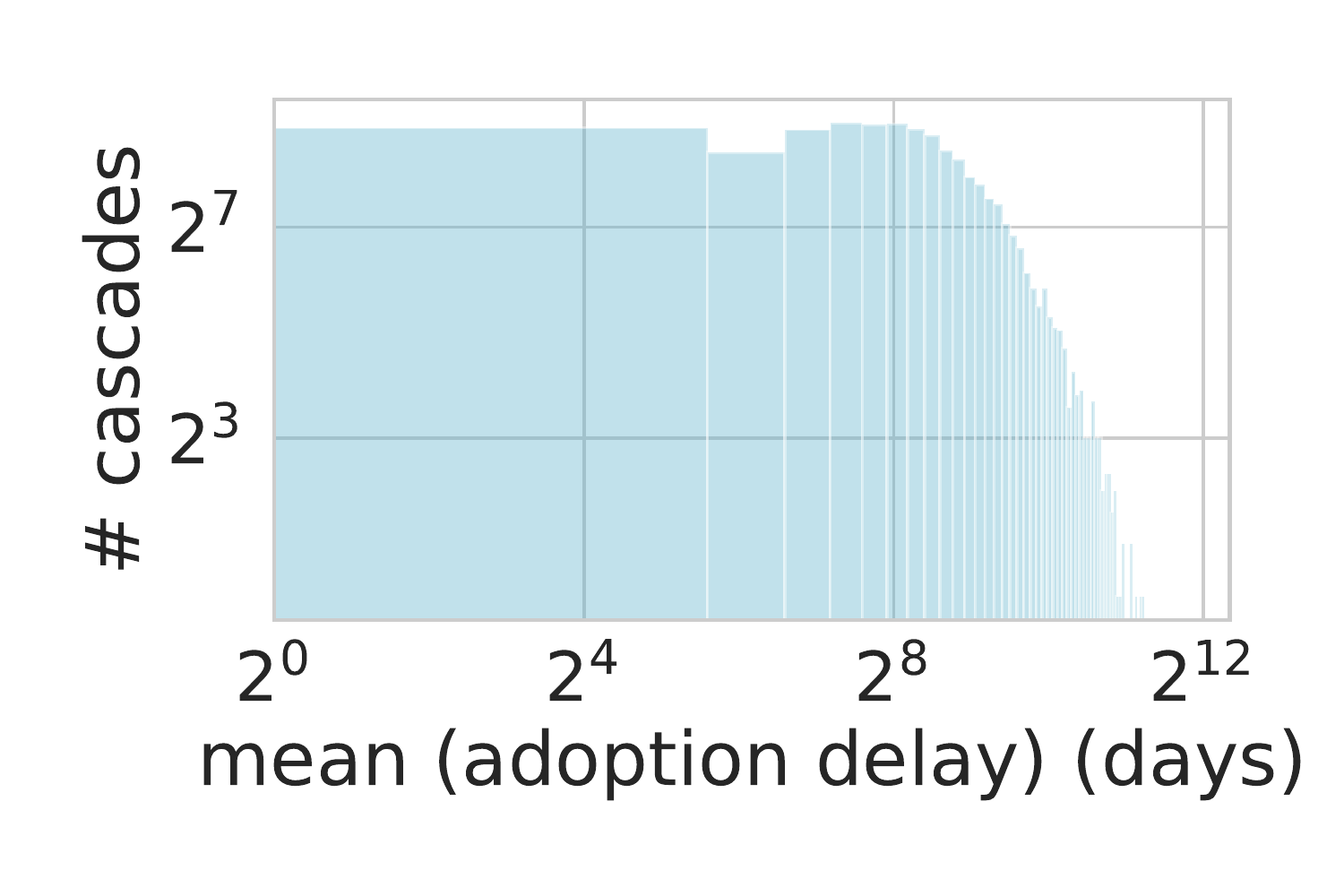}
			\label{fig:github-fork_cascade_delay}
	}
	\\
	
	\subfloat[Reddit]{
		\includegraphics[width=0.24\textwidth]{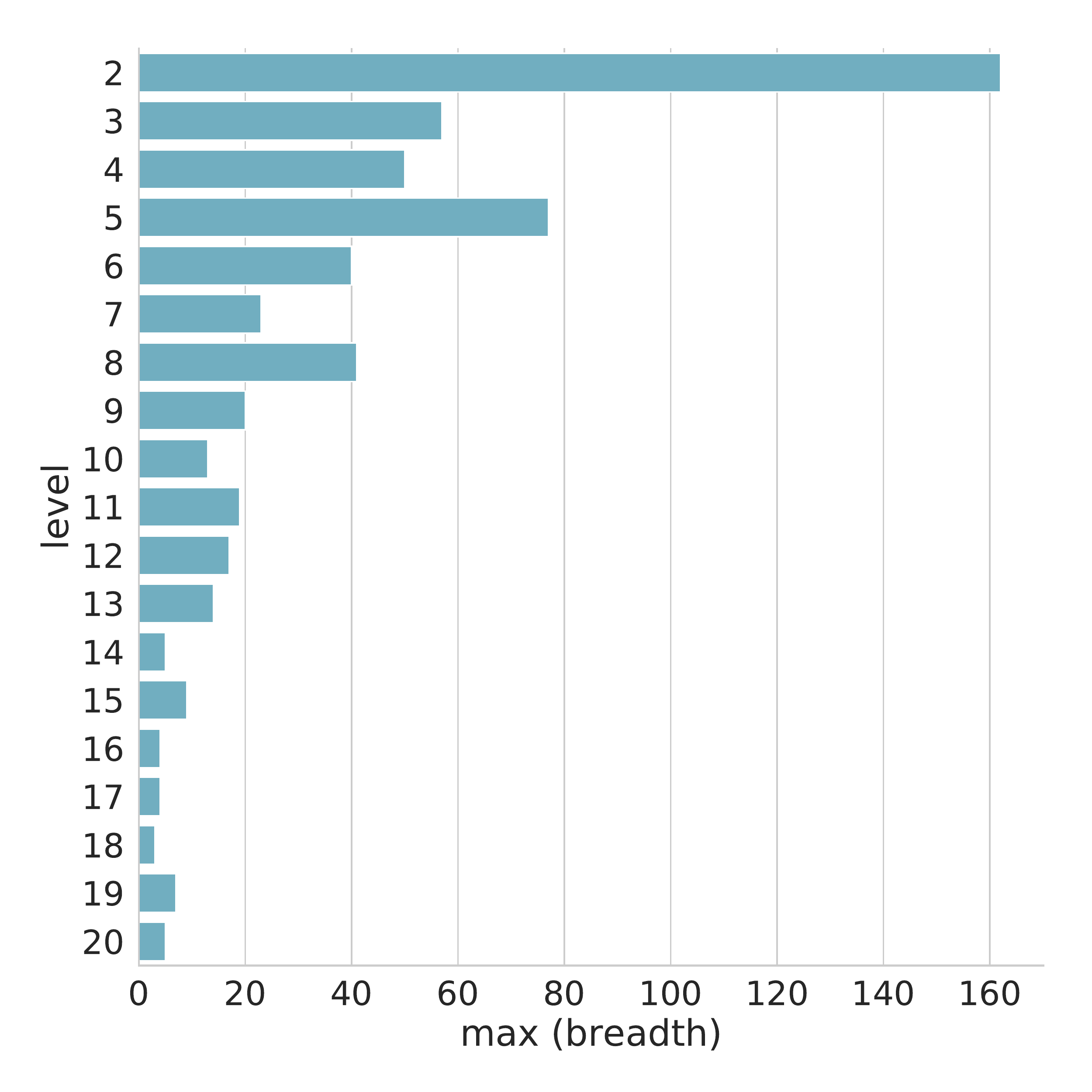}
			\label{fig:crypto_cascade_breadth_level}
	}
	
	&
		\subfloat[Github]{
		\includegraphics[width=0.24\textwidth]{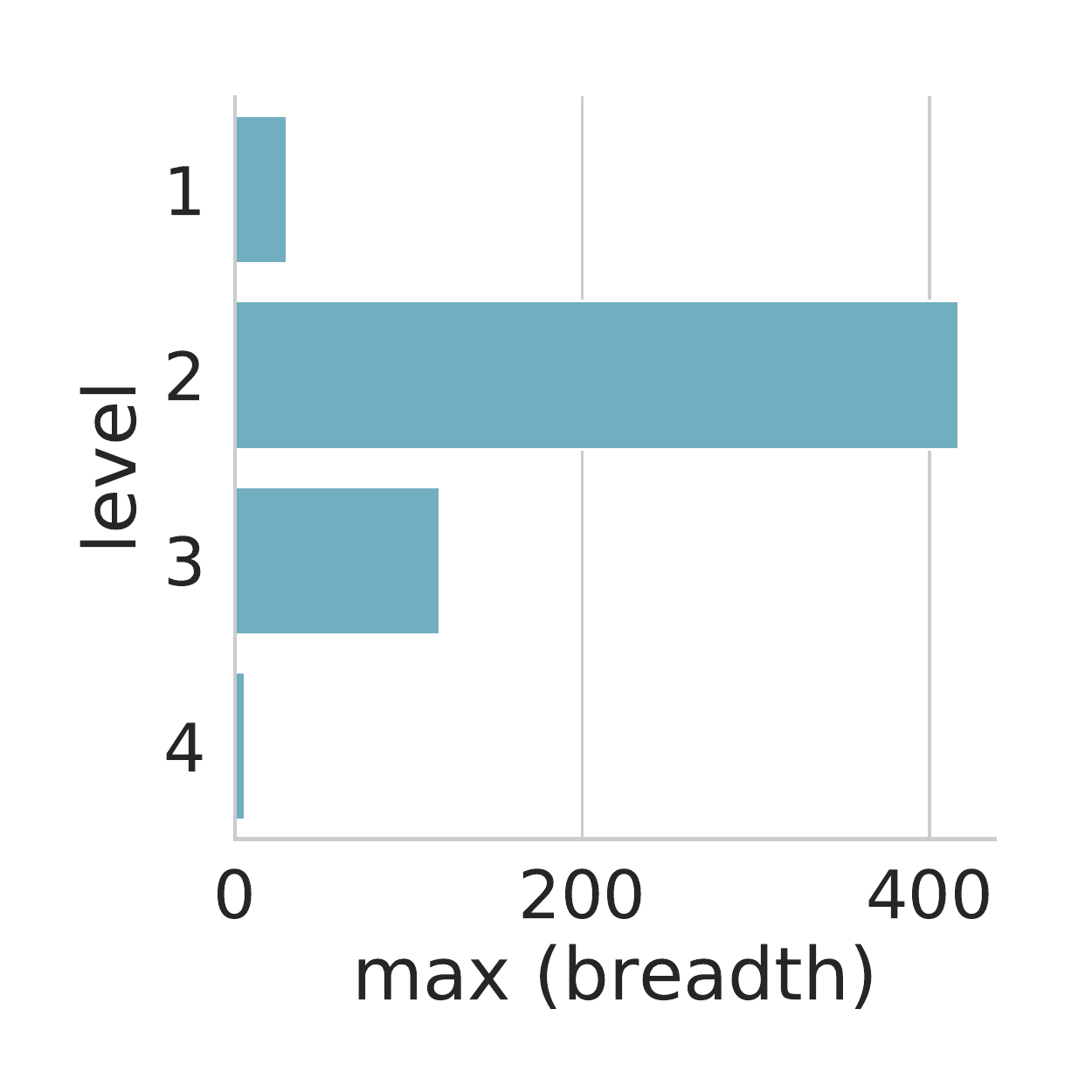}
			\label{fig:github-fork_cascade_breadth_level}
	}
	\\

\end{tabular}
	 \caption{A comparison of basic characteristics of cascades in Reddit and Github; the distribution of cascades is presented by the number of adoptions (size) in (a),(b), by the max depth in (c),(d), by the mean adoption delay in (e),(f) respectively, while (g), (h) present the max breadth by the level observed in the cascades. Plots in the same row compare the particular property over two platforms.}
	 \label{fig:cascade_chars}
	 
\end{figure}

\begin{figure}[htbp]
    \centering
    \includegraphics[width=0.5\textwidth]{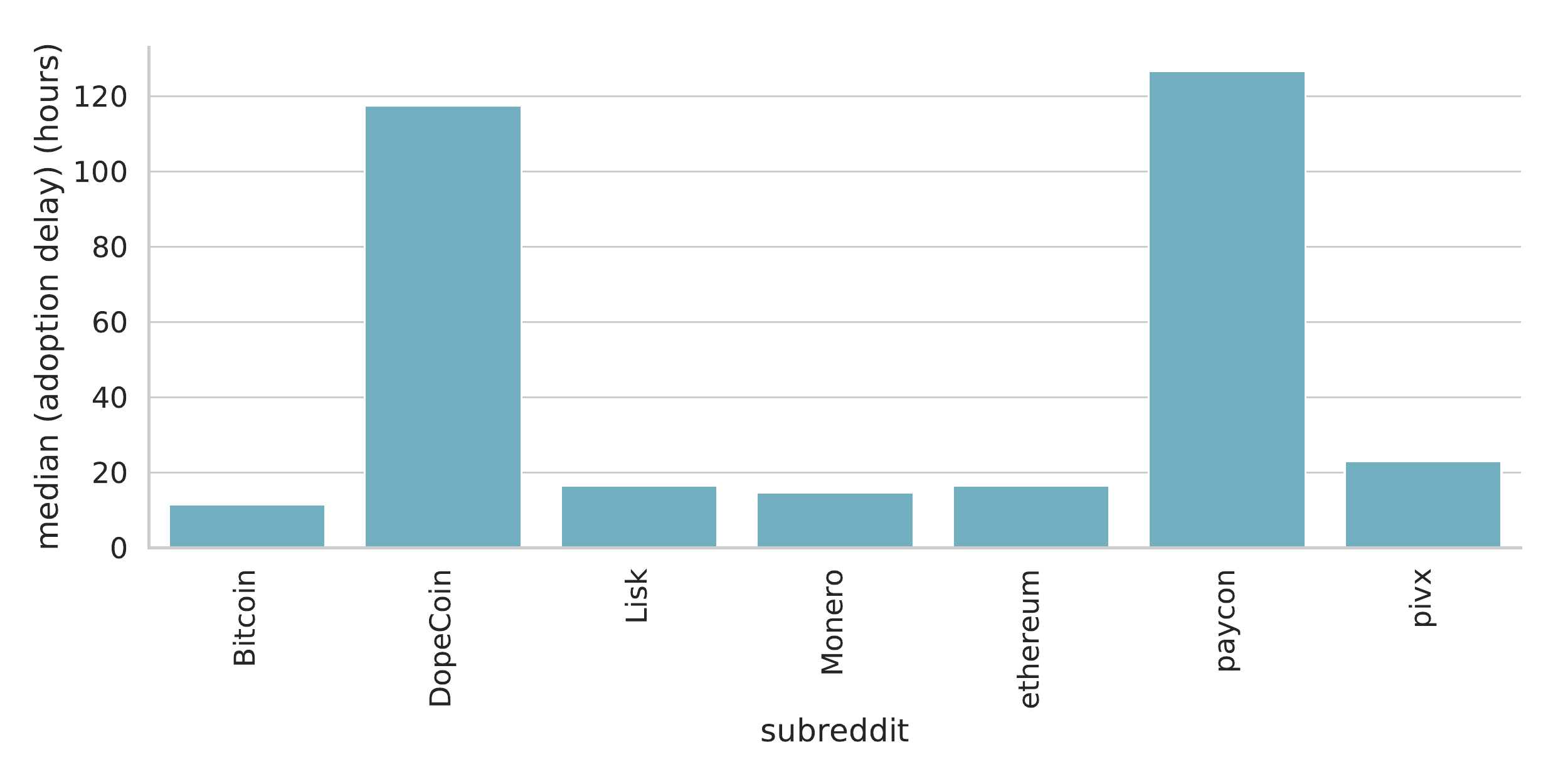}
    \caption{Reddit: Mean delay of adoptions observed in cascades categorized by subreddit.}
    \label{fig:cascade_delay_subreddit}
\end{figure}

The main description of the cascades in the two platforms we study are presented in Table~\ref{tbl:cascade_props}. 
We use four spatio-temporal properties to compare observed cascades in the two platforms: i) size, ii) max depth, iii) max breadth by level and iv) adoption delay.
Figure~\ref{fig:cascade_chars} presents such comparison.

\emph{Cascade Size:} We observe larger cascades in Reddit than Github as shown in Figures~\ref{fig:crypto_cascade_size} and ~\ref{fig:github-fork_cascade_size}. 
The mean cascade size in Reddit is 17 while in Github 6.
The largest cascades, as seen in Table~\ref{tbl:cascade_props}, are 7,868 for Reddit and 421 for Github.

We also fit the distribution of cascade sizes into a power law distribution, where $\alpha$ is the power law exponent, and obtained $\alpha=3.57$ for Reddit and $\alpha=2.63$ for Github.
(For a meaningful comparison, we removed the one-level depth cascades from Reddit in this case).

\emph{Maximum cascade depth:} Figures~\ref{fig:crypto_cascade_max_depth} and \ref{fig:github-fork_cascade_max_depth} show the distribution of cascades by maximum cascade depth.
In a discussion forum such as Reddit, it is common to find long conversation threads -- the mean cascade depth is 4.
We observe the reverse pattern in Github, where it is very rare to find long fork chains of Github repositories, and thus the mean depth is 2.

\emph{Maximum cascade breadth by level:} We observed a very interesting phenomena in Github cascades. As shown in Figure~\ref{fig:github-fork_cascade_max_depth}, the most attractive repository in the cascade is no longer the root of the cascade, but the immediate child of the root. 
(As a reminder, we only consider non-star cascades in Github). 
In Reddit conversations, the post is usually the most attractive component in the cascade: maximum breadth is at root level, 7,578.
However, we do not observe a consistent decrease of the attraction (as measured by breath of the cascade per level) when the level increases (Figure~\ref{fig:crypto_cascade_breadth_level}).

\emph{Adoption delay:} Github cascades are much slower in the pace of adoption than Reddit cascades, as shown in Figures~\ref{fig:crypto_cascade_delay} and ~\ref{fig:github-fork_cascade_delay}.
There are $67830\ (32\%)$ cascades in our Reddit dataset and only $96\ (1.6\%)$ cascades in Github with lifespan under a day. 

\begin{figure}[htbp]
    \centering
    \includegraphics[width=0.5\textwidth]{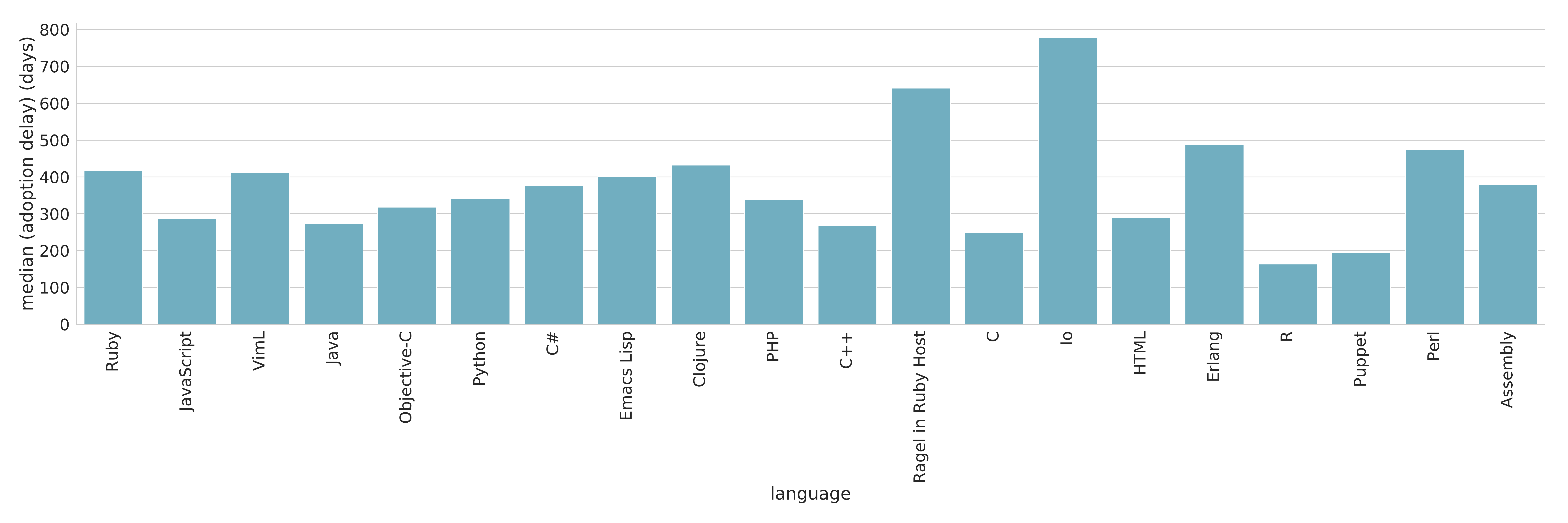}
    \caption{Github: Mean delay of adoptions observed in cascades categorized by the main programming language of the root repository. First 20 languages are visualized ordered by the date of appearance in the sampled dataset.}
    \label{fig:cascade_delay_lang}
\end{figure}

\emph{Community-level Analysis:} We also compare the delay of adoption at community level.
In Reddit, communities are explicitly defined by subreddits which, given our dataset selection, center around different crypto-currency systems, such as Bitcoin, Monero, Ethereum, etc.
In Github, we define a community by the programming language declared as the main language of a repository.
Figure~\ref{fig:cascade_delay_subreddit} shows the delay of adoption over Reddit communities (i.e., subreddits) while Figure~\ref{fig:cascade_delay_lang} presents the mean delay of adoptions for the Github repositories under a particular language.
In Reddit, we observe the popular crypto-currency systems like Bitcoin, Monero and Ethereum have fast rates of adoptions.
On the other hand, DopeCoin, which is especially made for cannabis enthusiasts, has a relatively slow rate of adoptions.
Private Instant Verified Transaction(PIVX) and Lisk(LSK) are new crypto-coins started trade in 2017 February, and have fast rate of adoption.
In Github, we only show the first 20 languages by the order of their appearances in our dataset. 
\emph{R} and Puppet are some languages to have an adoption delay under 200 days.

\subsection{Node Features}
\label{sec:datasets-node-features}

\begin{table*}[htbp]
    \centering
    
    \caption{Description of a node participates in the cascade. A node consists of an author and the attached comment/ post. Two binary target values are derived for the prediction task, i) whether a node participates in the spread or not ($node\_degree>0$) ii) whether a node has an early or late adoption relative to the other nodes participate in the cascade ($node\_short\_delay>median(the\ distribution\ of\ adoption\ delays\ in\ the\ cascade)$)}
    \resizebox{\linewidth}{!}{
    \begin{tabular}{|c|c|l|l|}
    \hline
    Feature Domain & Platform & Feature Name & Description \\ \hline \hline
    \multirow{8}{*}{Cascade Features} & \multirow{8}{*}{Both} & \textbf{$node\_degree$} & Number of comments for comment/post \\ \cline{3-4}
    & & \textbf{$node\_short\_delay$} & Adoption delay from the parent comment/post or repository \\ \cline{3-4}
    & & \textbf{$node\_long\_delay$} & Adoption delay from the root post/root repository \\ \cline{3-4}
         & & $node\_level$ & Level of the cascade tree \\ \cline{3-4}
         & &$node\_birth\_order$ & Birth order of comment/repository \\\cline{3-4}
         & & $p\_node\_degree$ & Number of comments for the parent comment/post or repository \\\cline{3-4}
         & & $p\_node\_birth\_order$ & Birth order of the parent comment/parent repository \\\cline{3-4} 
         & & $gp\_node\_degree$ & Number of comments to the grandparent comment/post or repository\\\cline{3-4}
       & & $gp\_node\_birth\_order$ & Birth order of the grandparent comment or repository \\\hline

        \multirow{11}{*}{User Features} & \multirow{3}{*}{Reddit} & $node\_author\_past\_no\_comments$ & Total number comments received by the comment author in the past \\ \cline{3-4}
       & & $node\_author\_past\_score$ & Total netscore (upvotes$-$downvotes) of the comment author in the past \\ \cline{3-4}
       & & $node\_author\_past\_no\_acts$ & Total number comments made by comment author in the past \\\cline{2-4}

        & \multirow{8}{*}{Github} & $node\_author\_age$ & Age of the author on the creation of given repository  \\ \cline{3-4}
        &  & $p\_node\_author\_age$ & Age of the parent author on the creation of parent repository  \\ \cline{3-4}
        &  & $gp\_node\_author\_age$ & Age of the grand parent author on the creation of grand parent repository  \\ \cline{3-4}
        &  & $root\_node\_author\_age$ & Age of the root author on the creation of root repository  \\ \cline{3-4}
       & & $node\_author\_influence\_score$ & Influence score of the author based upon the stars received \\ \cline{3-4}
       & & $node\_author\_public\_repos$ & Total number of public repositories owned by the author \\ \cline{3-4}
       & & $node\_author\_no\_followers$ & Number of followers of the author \\ \cline{3-4}
       & & $node\_author\_no\_following$ & Number of users the author follows \\ \hline
       
       \multirow{15}{*}{Content Features} & \multirow{9}{*}{Reddit} & $comment\_score$ & Netscore of the comment \\\cline{3-4}
       & & $comment\_subjectivity$ & Subjectivity score of the comment \\\cline{3-4}
       & & $comment\_controversiality$ & Controversiality score of the comment \\ \cline{3-4}
       & & $p\_comment\_score$ & Netscore of the parent comment \\ \cline{3-4}
       & & $p\_comment\_subjectivity$ & Subjectivity score of the parent comment/ post \\ \cline{3-4}
       & & $p\_comment\_controversiality$ & Ccontroversiality score of the parent comment \\ \cline{3-4}
       & & $gp\_comment\_score$ & Netscore of the grand parent comment \\ \cline{3-4}
       & & $gp\_comment\_subjectivity$ & Subjectivity score of the grand parent comment/ post \\ \cline{3-4}
       & & $gp\_comment\_controversiality$ & Controversiality score of the grand parent comment \\ \cline{2-4}
       
       & \multirow{8}{*}{Github} & $repo\_open\_issue\_count$ & Number of open issues in the repository \\\cline{3-4}
       & & $repo\_no\_watchers$ & Number of watchers of the repository \\\cline{3-4}
       & & $p\_repo\_open\_issue\_count$ & Number of open issues in the parent repository \\\cline{3-4}
       & & $p\_repo\_no\_watchers$ & Number of watchers of the parent repository \\\cline{3-4}
       & & $gp\_repo\_open\_issue\_count$ & Number of open issues in the grand parent repository \\\cline{3-4}
       & & $gp\_repo\_no\_watchers$ & Number of watchers of the grand parent repository \\\cline{3-4}
       
       & & $root\_repo\_open\_issue\_count$ & Number of open issues in the root repository \\\cline{3-4}
       & & $root\_repo\_no\_watchers$ & Number of watchers of the root repository \\\hline
    \end{tabular}
    }
    \label{tbl:feature_desc}
\end{table*}

Table~\ref{tbl:feature_desc} presents all the features that we used to describe a node (original post or comment) that participates in a cascade.
We categorize these features into three groups: i)~spatio-temporal properties, ii)~user-level properties, and iii)~content-level properties.

\subsubsection{Spatio-temporal Properties}
We represent the topology around an individual node in the cascade using two spatio-temporal properties: degree (i.e., number of adoptions) and the birth order of the predecessors.
As an example, we use the degree and birth order of the parent (level $i-1$) and the grand-parent (level $i-2$) nodes to represent a node in level $i$.

\subsubsection{User-level Features}
In both Reddit and Github, users control the formation of  cascades by posting comments or forking repositories.
These actions could be in response to the users who authored the previous messages/repositories rather than simply to the content users interact with. 
We thus represent a Reddit user via a set of features describing her importance in the platform. 
In our setting, the importance of a Reddit user is measured by the amount of activity she has done prior to the particular reaction (see Table~\ref{tbl:feature_desc}).

We calculate the age of a Github user account 
at the time of a particular action.
The age values of the predecessors used to describe a node in a given level.
We also extract the influence score (i.e., gh-impact) of a Github user which is calculated based upon the stars received for her projects (an implementation of gh-impact is used to extract this feature~\cite{gh-impact}).
Table~\ref{tbl:feature_desc} describes other features used to describe the Github user.

\subsubsection{Content-level Features}
As we described earlier, Reddit conversations contain more textual content.
An individual might prefer the content diffused in the cascade to perform a reaction.
We extract the sentiment scores of Reddit comment that quantify the subjective, and controversial content (a Python library of a natural language toolkit is used to calculate this feature~\cite{bird2004nltk}).
We also capture the semantic structure of the comments at predecessor nodes (see Table~\ref{tbl:feature_desc}).

In Github, a repository defines the content.
We extract two key features that explain the code-base (i.e., number of open issues) and the popularity of a repository (i.e., number of watches).
The repository features of the predecessors used to represent a node in a given level.

	
	
	
	 
\section{Predicting the Topological Structure of Cascades over Time}
\label{sec:results}

We present our results on the following objectives. 
First, we want to predict if a message in the cascade ends the discussion / fork branch or generates more responses. 
Second, we want to predict how quickly a comment (or repository) is made in response to the parent post/ comment (or parent repository) (Section~\ref{sec:temp_position}).

Building on these two components, we achieve our final objective of predicting the structure of the cascade over time (Section~\ref{sec:tree-shape}). 
We compare the performance of our generative approach with the set of baseline models presented. 
As we described in Section~\ref{sec:cascade-generation}, we use the conditional probability distribution of degree to construct the baseline cascade trees.

Table~\ref{tbl:stat_eval_data} presents the basic statistics of the training and testing data that we used in our experiments.
Reddit cascades are split by the post creation time, with training data between January 2015 and December 2016 and testing data from January to August 2017. Github cascades are split randomly by language, such that the ratio training to testing size is 2 to 1.

\begin{table}[htbp]
	\centering
	\caption{Basic statistics of training/testing datasets. 
	}
	\label{tbl:stat_eval_data}
	\begin{tabular}{|r|r|r|r|r|}
		\hline 
		\multirow{2}{*}{Platform} & \multicolumn{2}{|c|}{Training} & \multicolumn{2}{|c|}{Testing}
\\		\cline{2-5}
		& $\#$ cascades &  $\#$ nodes & $\#$ cascades &  $\#$ nodes \\ 
		\hline 
		
		
		Reddit & 128,359 & 2.3M & 81,362 & 1.3M \\ \hline
		
		Github & 4,048 & 25K & 2,024 & 13K \\ \hline
		
		
		\end{tabular} 
\end{table}

\subsection{Predicting Temporal Position in the Message Tree}
\label{sec:temp_position}

\begin{table*}[bthp]
	\centering
	\caption{Model performance on unseen cascade records in the testing period. Results are shown for two units of target values, i) degree ii) speed of individual adoptions, and they are compared with several baseline models. Two empirically bench-marked conditional probabilistic distributions are used to derive the target values in the baseline models. The level and the birth order of a node participates in a cascade are two such conditional attributes (presented inside brackets).
	}
	\label{tbl:results}
	\begin{tabular}{|r|r|r|r|r|r|r|r|r|r|}
		\hline 
	\multirow{2}{*}{Target Unit} & \multirow{2}{*}{Platform} & \multirow{2}{*}{Model} & \multirow{2}{*}{Accuracy} &  \multicolumn{2}{|c|}{Precision} & \multicolumn{2}{|c|}{Recall} &
		\multicolumn{2}{|c|}{F1-score} \\
		\cline{5-10}
		& & & & class 0 & class 1 &  class 0 & class 1 & class 0 & class 1 \\ \hline \hline
		
		
		

		\multirow{6}{*}{Branch or Leaf Node} & \multirow{3}{*}{Reddit} & Baseline (degree, level) & 0.54 & 0.52 & 0.55 & 0.52 & 0.54 & 0.52 & 0.55 \\ \cline{3-10}
        
        & & Baseline (degree, birth order) & 0.57 & 0.44 & 0.49 & 0.52 & 0.53 & 0.48 & 0.55 \\ \cline{3-10}
		
		& & \textbf{Cascade-LSTM} & \textbf{0.73} & \textbf{0.70} & \textbf{0.75} & \textbf{0.73} & \textbf{0.72} & \textbf{0.71} & \textbf{0.74} \\ \cline{2-10}

			& \multirow{3}{*}{Github} & Baseline (degree, level) & 0.90 & 0.93 & 0.82 & 0.93 & 0.83 & 0.93 & 0.82 \\ \cline{3-10}
        
        & & Baseline (degree, birth order) & 0.70 & 0.93 & 0.12 & 0.73 & 0.43 & 0.82 & 0.20 \\ \cline{3-10}
		
		& & \textbf{Cascade-LSTM} & \textbf{0.98} & \textbf{0.98} & \textbf{0.97} & \textbf{0.99} & \textbf{0.96} & \textbf{0.98} & \textbf{0.97} \\ \hline \hline


    \multirow{6}{*}{Early or Late Adopter} & \multirow{3}{*}{Reddit} & Baseline (speed, level) & 0.55 & 0.56 & 0.54 & 0.57 & 0.54 &  0.57 & 0.54 \\ \cline{3-10}
        
        & & Baseline (speed, birth order) & 0.55 & 0.57 & 0.54 & 0.57 & 0.54 & 0.57 & 0.54 \\ \cline{3-10}
		
		& & \textbf{Cascade-LSTM} & \textbf{0.83} & \textbf{0.85} & \textbf{0.81} & \textbf{0.83} & \textbf{0.84} & \textbf{0.84} & \textbf{0.82} \\ \cline{2-10}

		& \multirow{3}{*}{Github} & Baseline (speed, level) & 0.59 & 0.62 & 0.57 & 0.62 & 0.56 & 0.62 & 0.56 \\ \cline{3-10}
        
        & & Baseline (speed, birth order) & 0.51 & 0.54 & 0.47 & 0.54 & 0.47 & 0.54 & 0.47 \\ \cline{3-10}
		
		& & \textbf{Cascade-LSTM} & \textbf{0.89} & \textbf{0.91} & \textbf{0.88} & \textbf{0.89} & \textbf{0.89} & \textbf{0.90} & \textbf{0.88} \\ \hline
		\end{tabular}
\end{table*}

We define two prediction tasks on the temporal position of a message in a cascade tree as described in Section~\ref{sec:pred_tasks}.

In the first prediction task (branch or leaf node), the two classes are balanced in the ratio of 65\%:35\% in Reddit and 70\%:30\% in Github.
In the next task on classifying early and late adopters, classes are perfectly balanced in both datasets.
 
In order to evaluate our LSTM-based prediction technique for these two prediction tasks, we generate baseline scenarios based on two probability distributions as follows.
For every node in the training dataset, we represent its level in the cascade tree, its degree, its birth order among its siblings, and its speed (0 or 1 for early or late adopter). 
Given a node in the testing dataset, we draw a degree, respectively a speed of adoption, with a weighted probability. 
In both cases we control by the level in the tree and, respectively, by the birth order, resulting in four independent baseline scenarios, two for each prediction task.  
The resulting baseline models are called Baseline (degree, level) and Baseline (degree, birth order) in Table~\ref{tbl:results} for predicting leaves/branch nodes, and Baseline (speed, level) and Baseline (speed, birth order) for predicting the pace of adoption.



We report the performance of our model and that of the baseline models in Table~\ref{tbl:results}.
Cascade-LSTM achieves $73\%$ accuracy in discriminating leaves vs. branching nodes in Reddit, and $98\%$ accuracy in Github.
In Reddit, the baseline models perform poorly, with $57\%$ accuracy, compared to Github, where accuracy is $90\%$.
This difference in the accuracy of the baseline models for the two platforms may be due to the radical difference between the cascade depths distributions: while in Reddit there is huge variation in depth, Github cascades are very shallow, with only three values for depth. 
The shape of a tall tree is much harder to predict, as it has more room for variation, compared to the shape of a very shallow tree.

The accuracy of our LSTM technique in distinguishing between early and late adopters is 83\% in Reddit and 89\% in Github, while the baseline models hover between 51\% and 59\% (which is slightly higher than expected for a random draw given the perfectly balanced classes). 

\begin{figure}[tbp]
	\centering
\begin{tabular}{cc}
	
	\subfloat[Baseline (level)]{
		\includegraphics[width=0.24\textwidth]{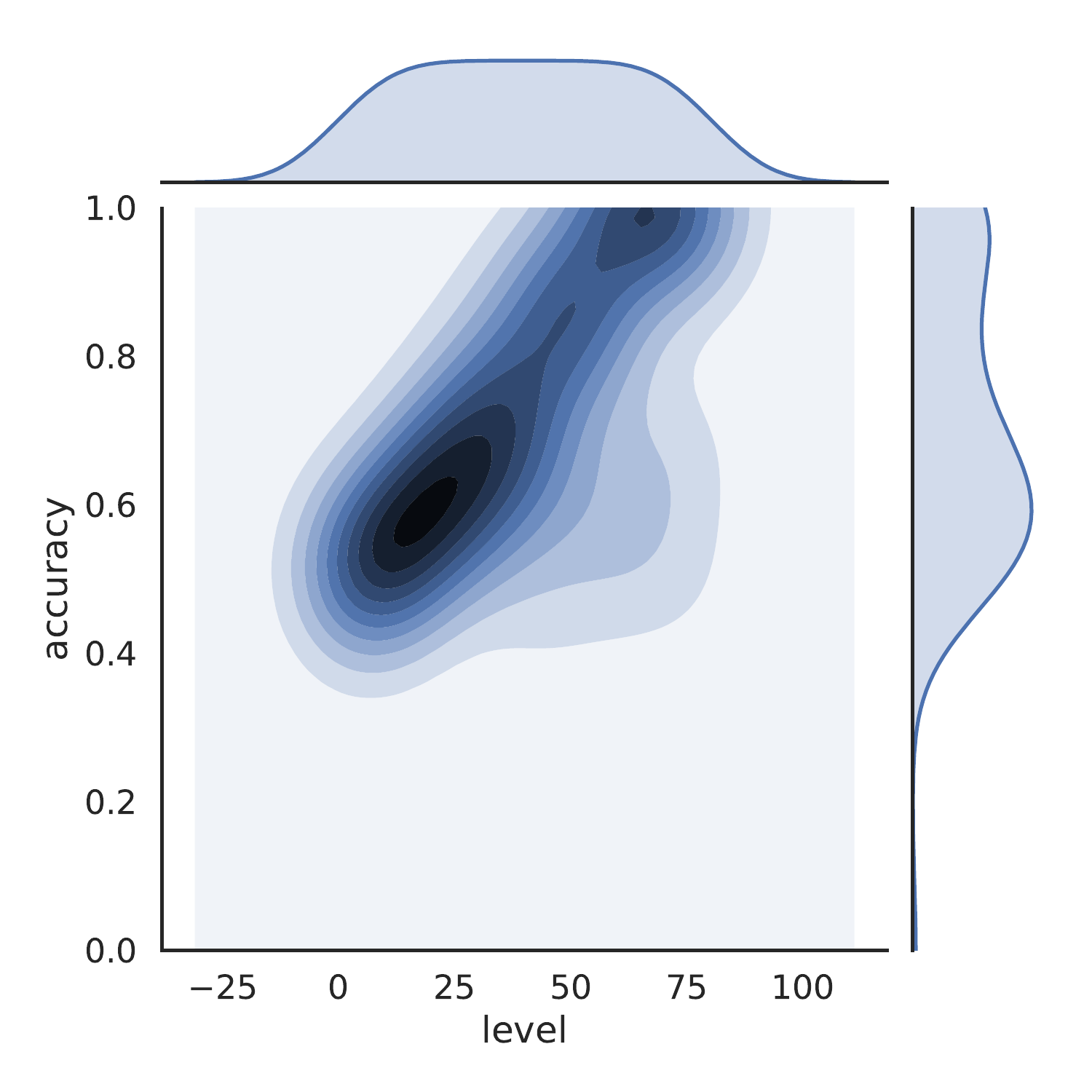}
			\label{fig:crypto_degree_accuracy_by_level_pmodel}
	}
	
&

	\subfloat[Cascade-LSTM]{
		\includegraphics[width=0.24\textwidth]{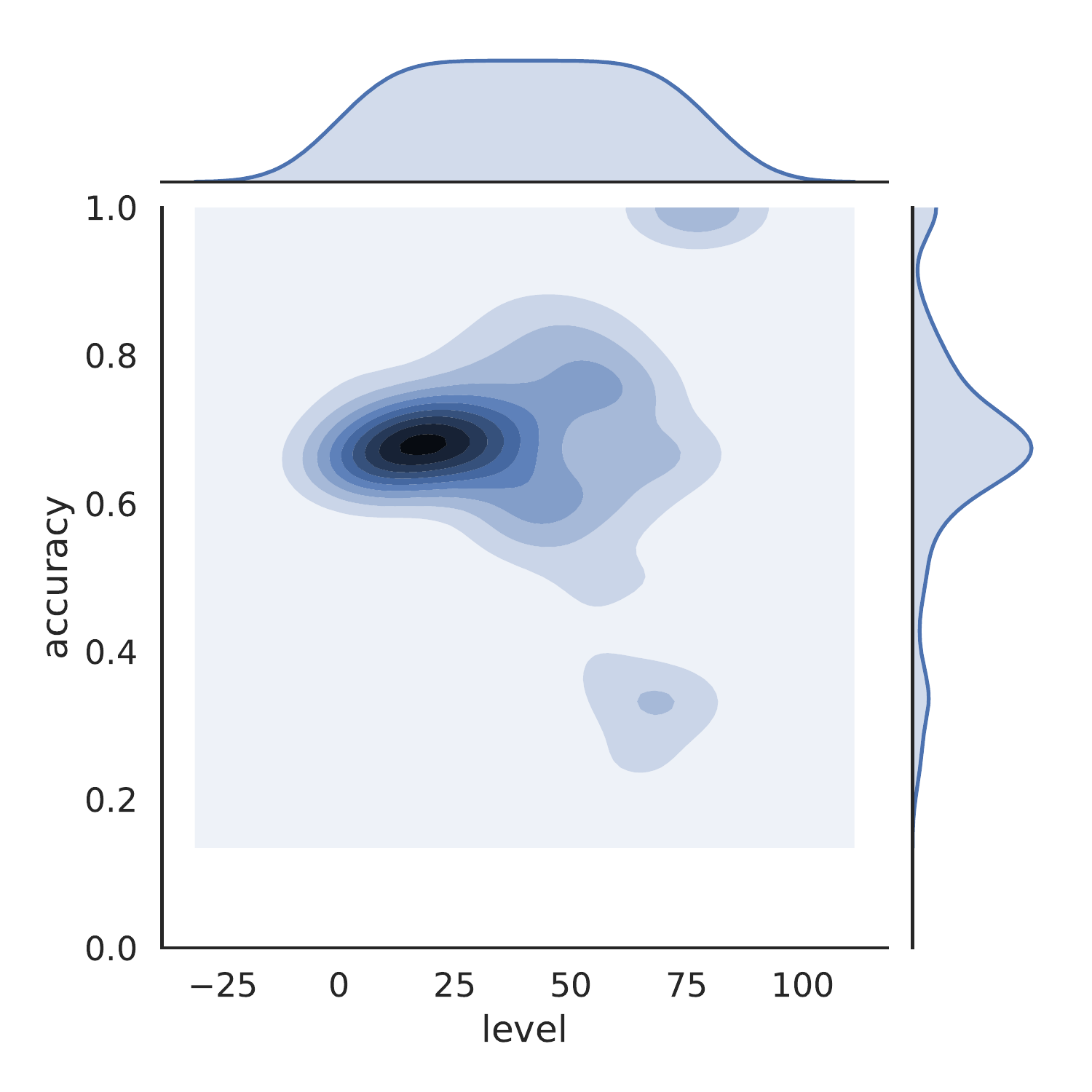}
			\label{fig:crypto_degree_accuracy_by_level_clstm}
	} 
	
	\\
	
		\subfloat[Baseline (birth order)]{
		\includegraphics[width=0.24\textwidth]{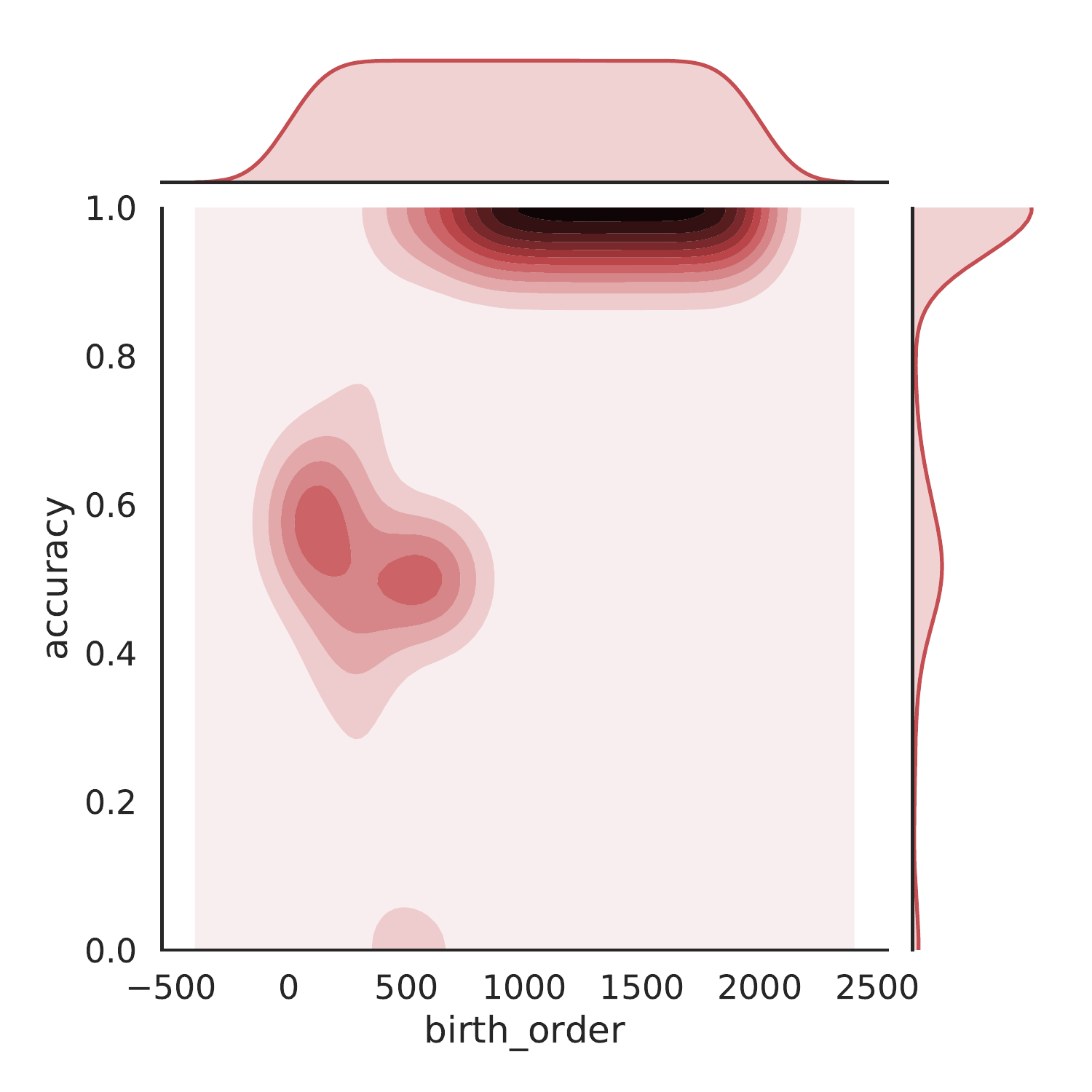}
			\label{fig:crypto_degree_accuracy_by_birth_order_pmodel}
	}
	
&

	\subfloat[Cascade-LSTM]{
		\includegraphics[width=0.24\textwidth]{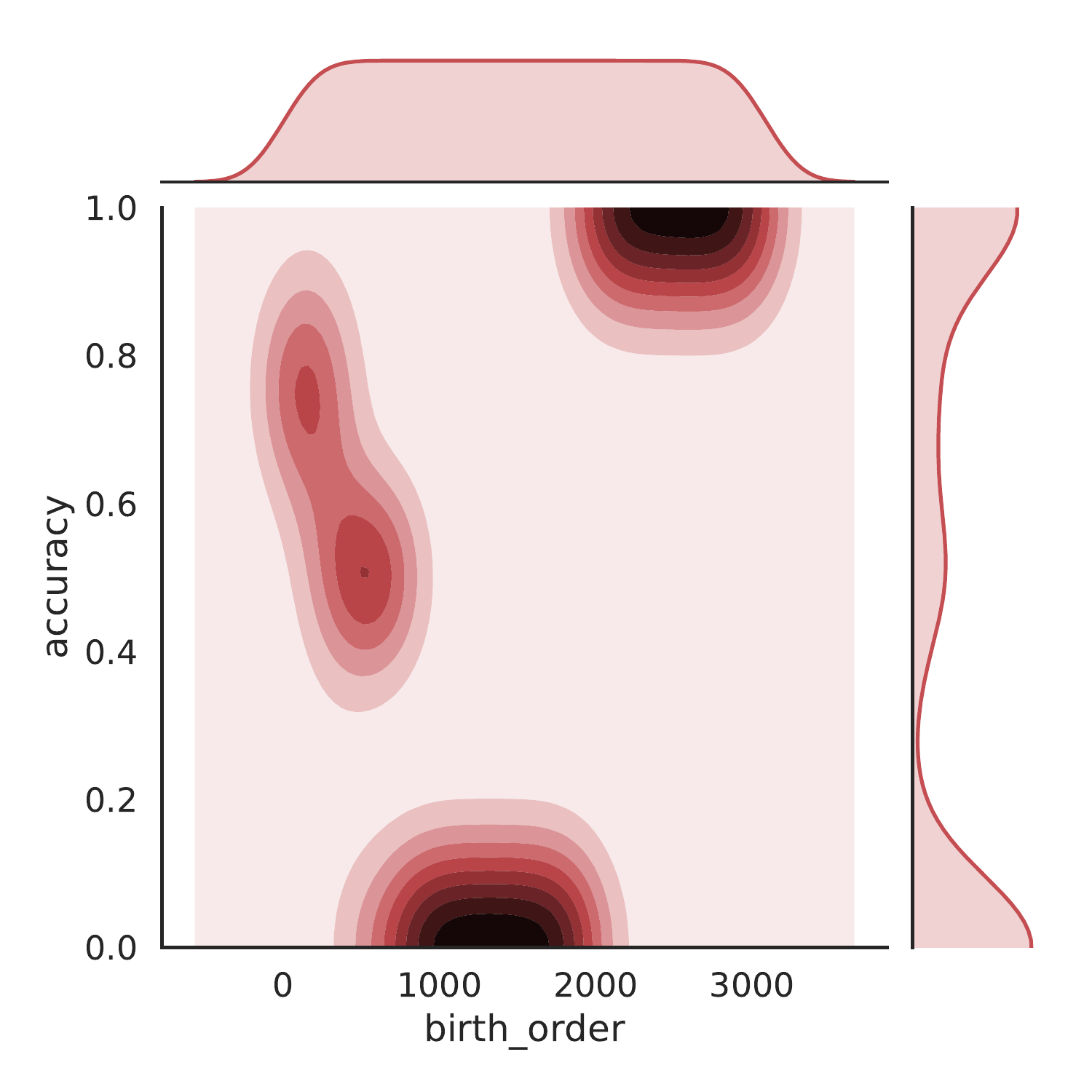}
			\label{fig:crypto_degree_accuracy_by_birth_order_clstm}
	}
	
\end{tabular}
	 \caption{Platform: Reddit, Accuracy Values of Degree conditioned by the level and birth order}
	 \label{fig:degree_accuracy_reddit}
	 
\end{figure}

\begin{figure}[t]
	\centering
\begin{tabular}{cc}

	\subfloat[Baseline (level)]{
		\includegraphics[width=0.24\textwidth]{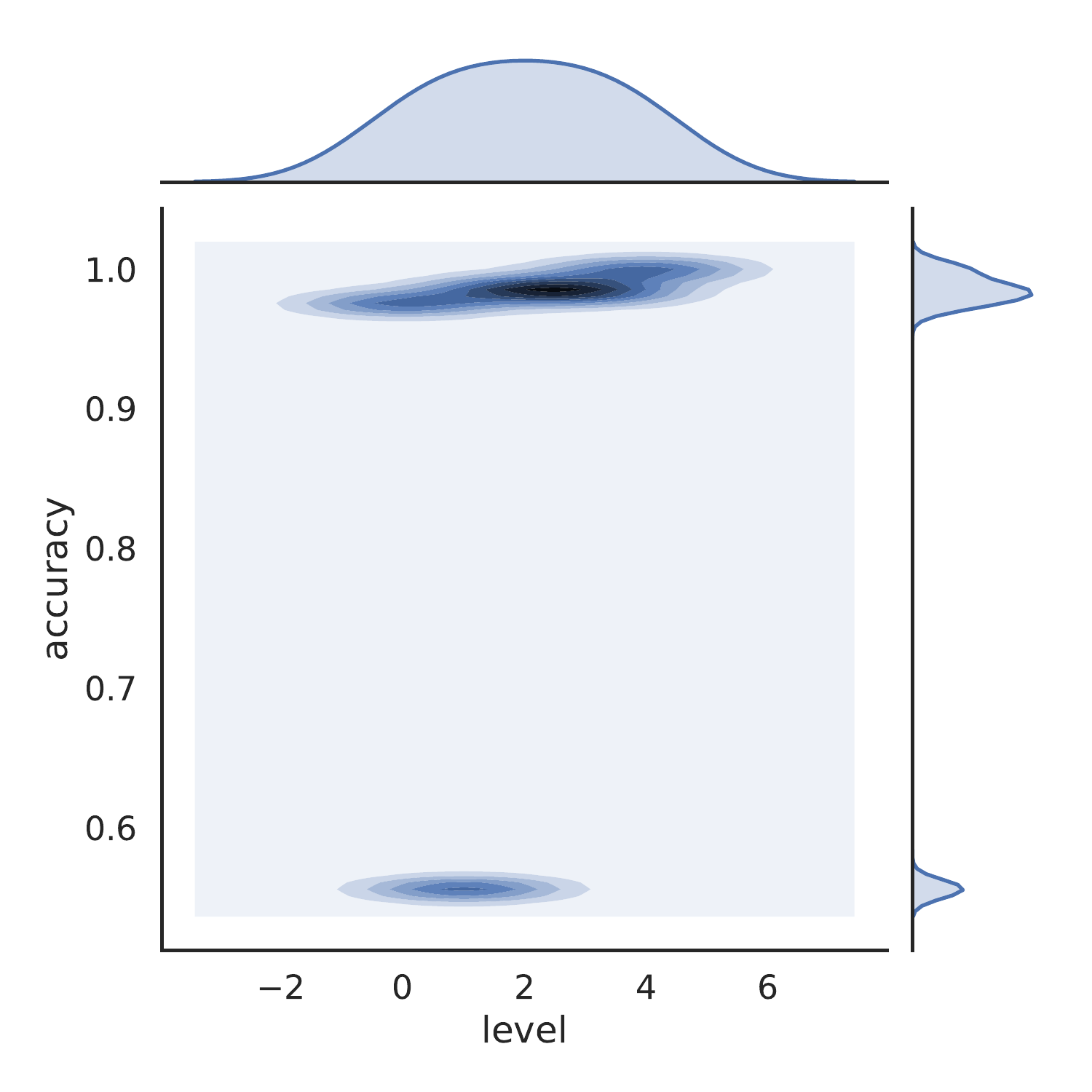}
			\label{fig:github_degree_accuracy_by_level_pmodel}
	}
	
&

	\subfloat[Cascade-LSTM]{
		\includegraphics[width=0.24\textwidth]{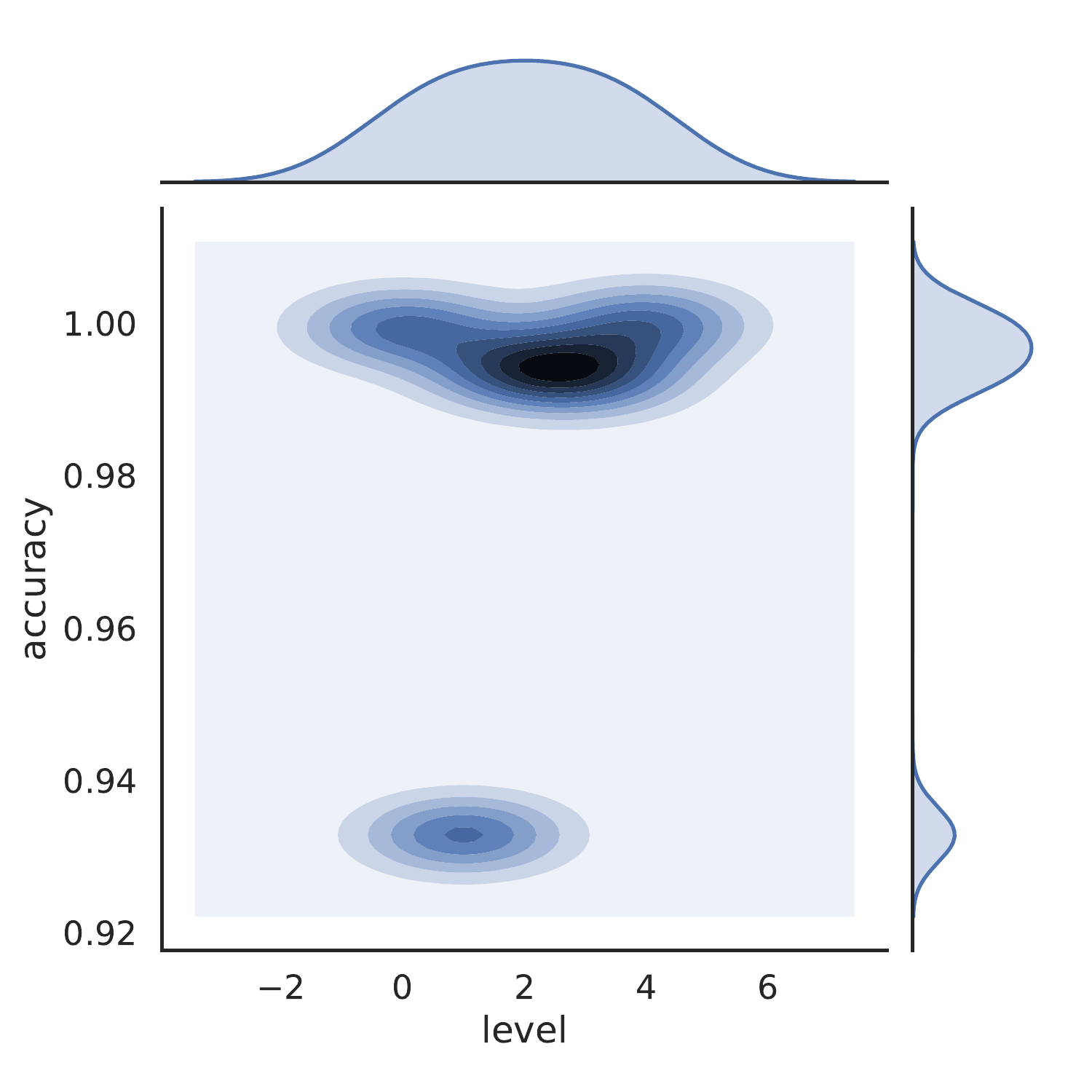}
			\label{fig:github_degree_accuracy_by_level_clstm}
	} 
	
	\\
	
		\subfloat[Baseline (birth order)]{
		\includegraphics[width=0.24\textwidth]{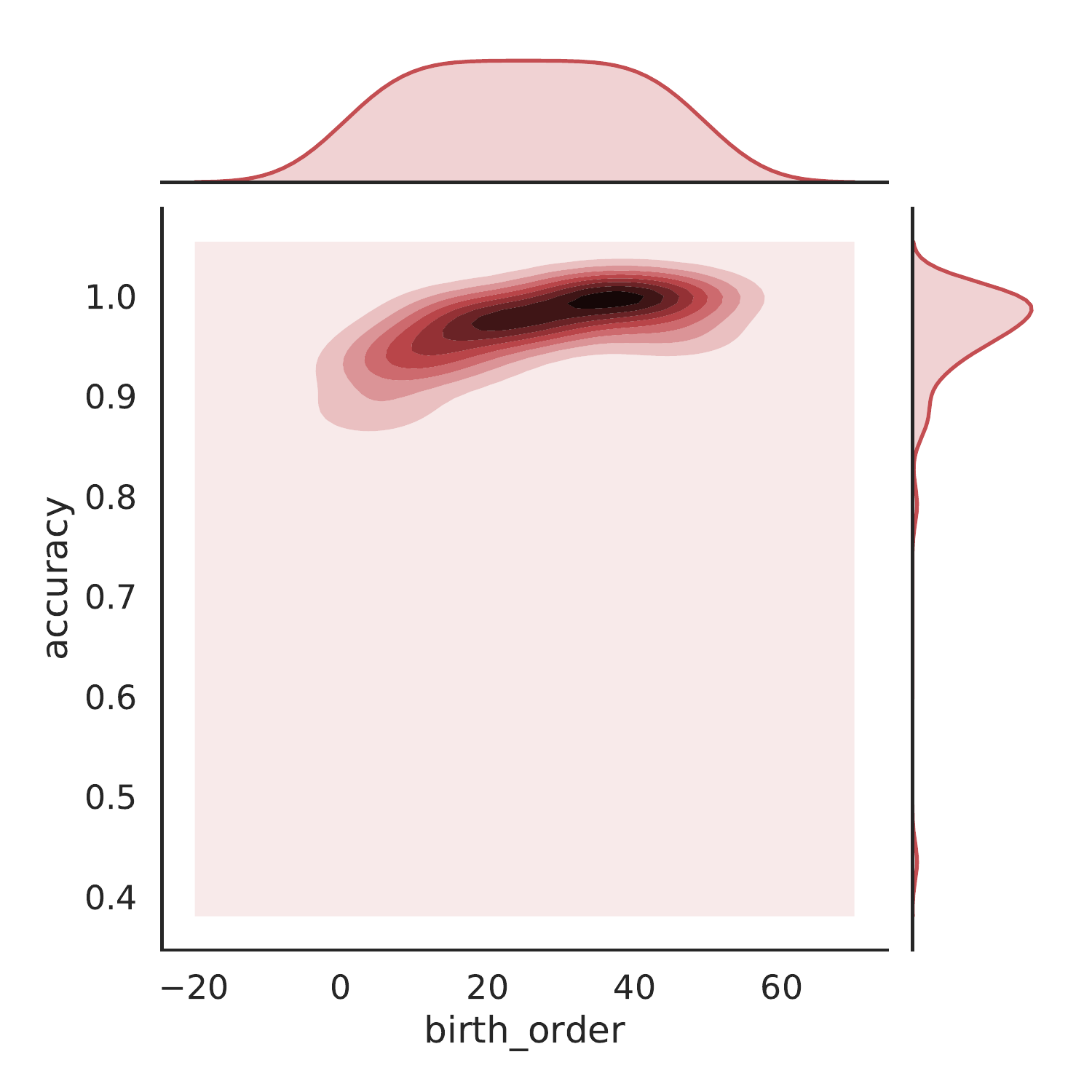}
			\label{fig:github_degree_accuracy_by_birth_order_pmodel}
	}
	
&

	\subfloat[Cascade-LSTM]{
		\includegraphics[width=0.24\textwidth]{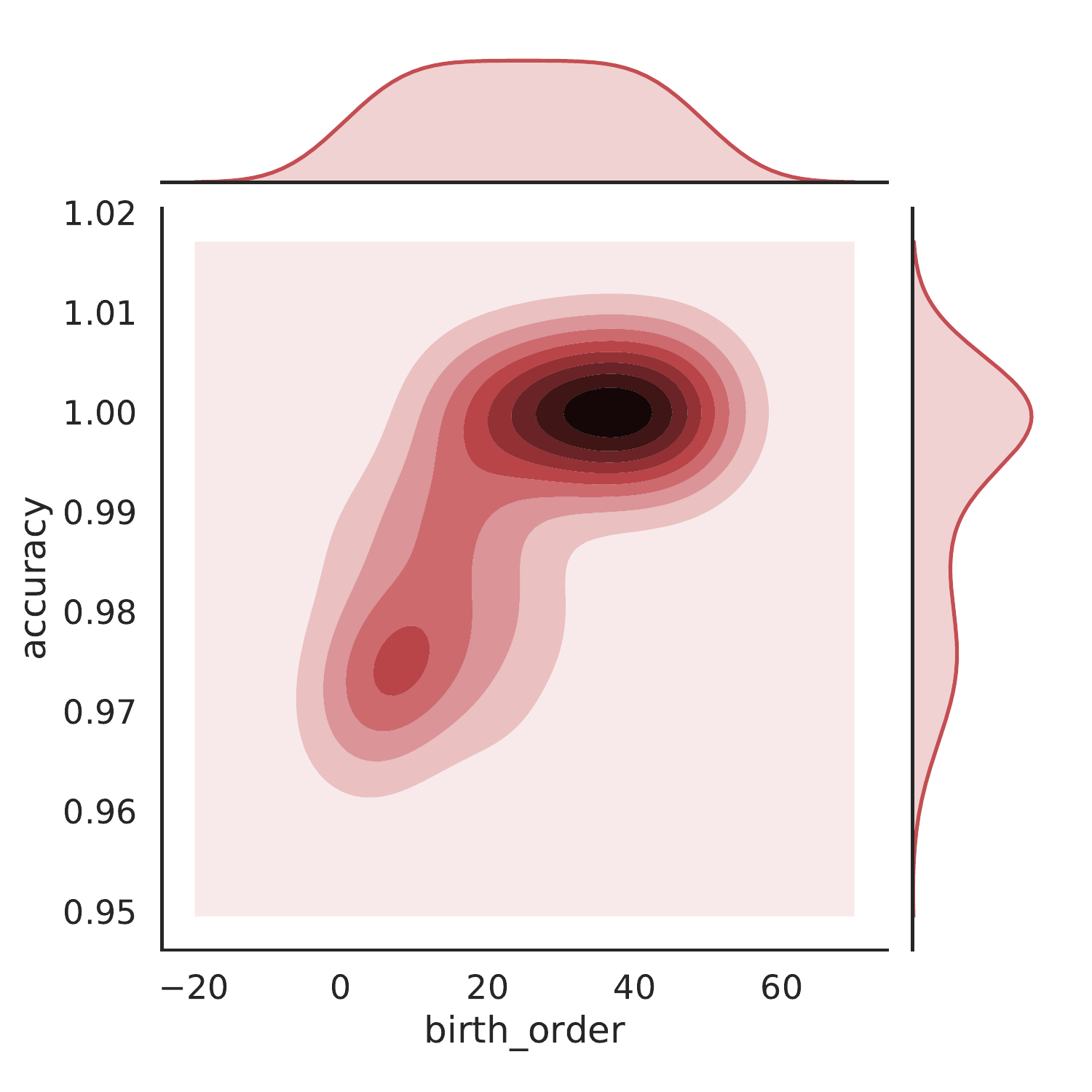}
			\label{fig:github_degree_accuracy_by_birth_order_clstm}
	}
	
\end{tabular}
	 \caption{Platform: Github, Accuracy Values of Degree conditioned by the level and birth order}
	 \label{fig:degree_accuracy_github}
	 
\end{figure}

\begin{figure}[t]
	\centering
	\centering
\begin{tabular}{cc}
	\subfloat[Baseline (level)]{
		\includegraphics[width=0.24\textwidth]{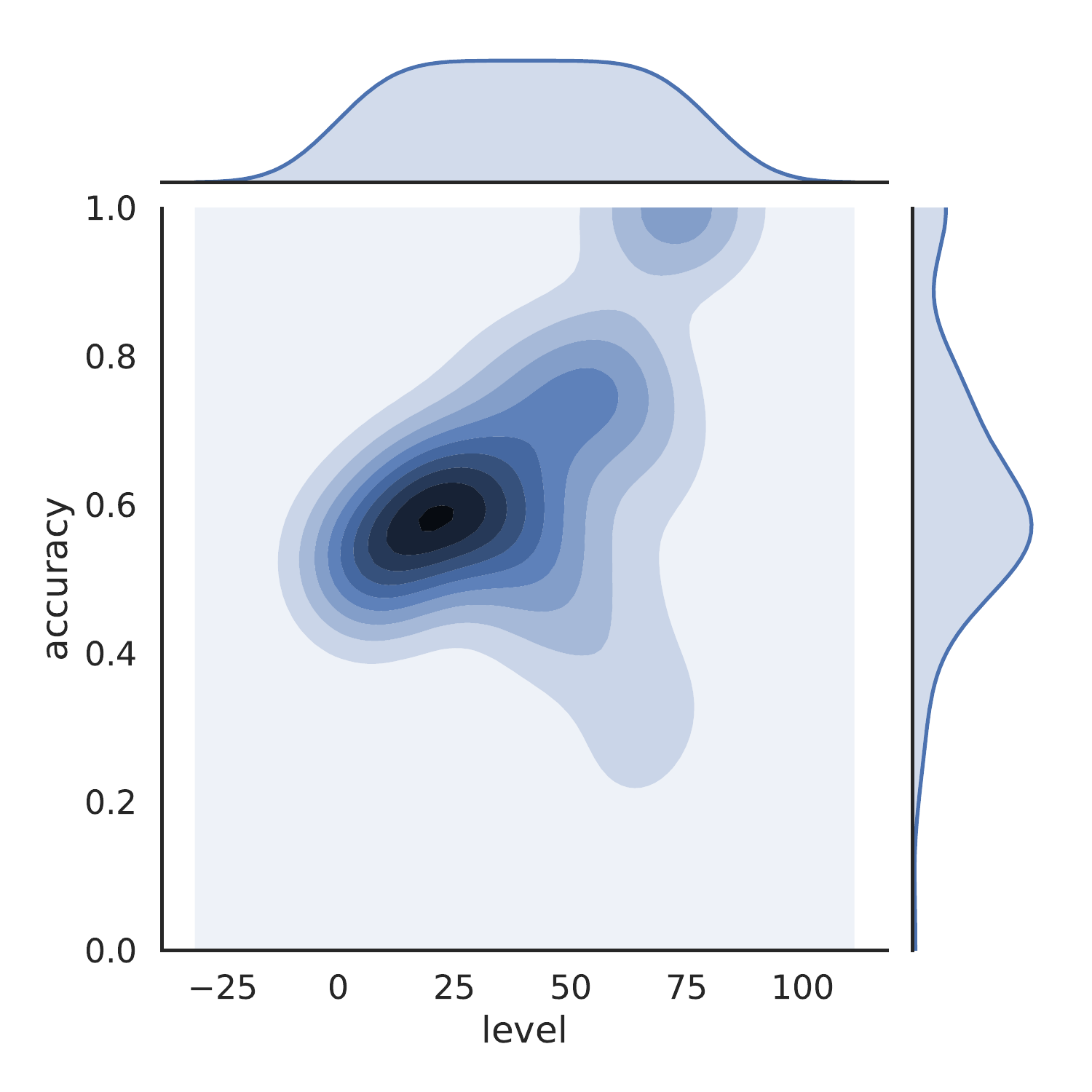}
			\label{fig:crypto_delay_accuracy_by_level_pmodel}
	}
	
&

	\subfloat[Cascade-LSTM]{
		\includegraphics[width=0.24\textwidth]{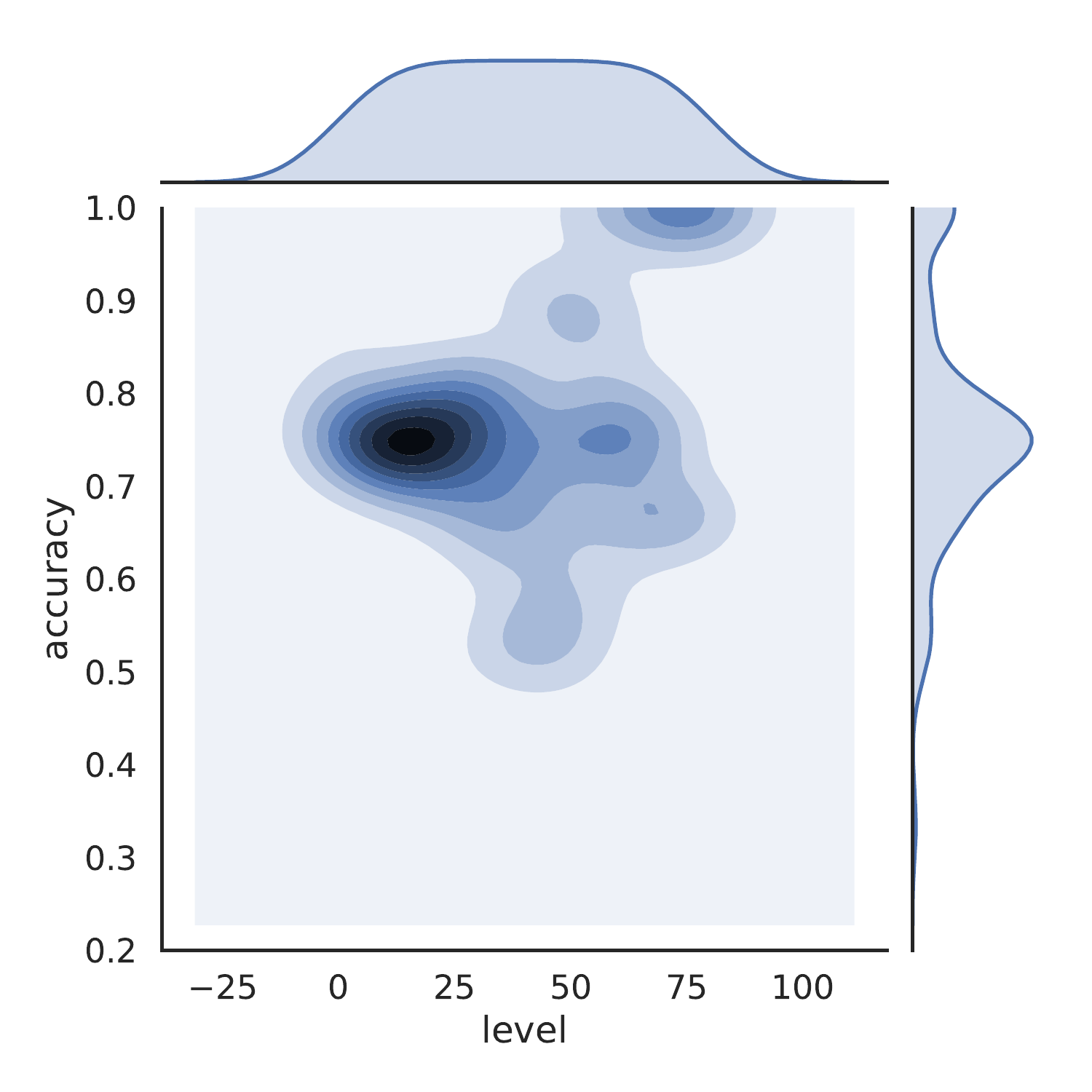}
			\label{fig:crypto_delay_accuracy_by_level_clstm}
	} 
	
	\\
	
	\subfloat[Baseline (birth order)]{
		\includegraphics[width=0.24\textwidth]{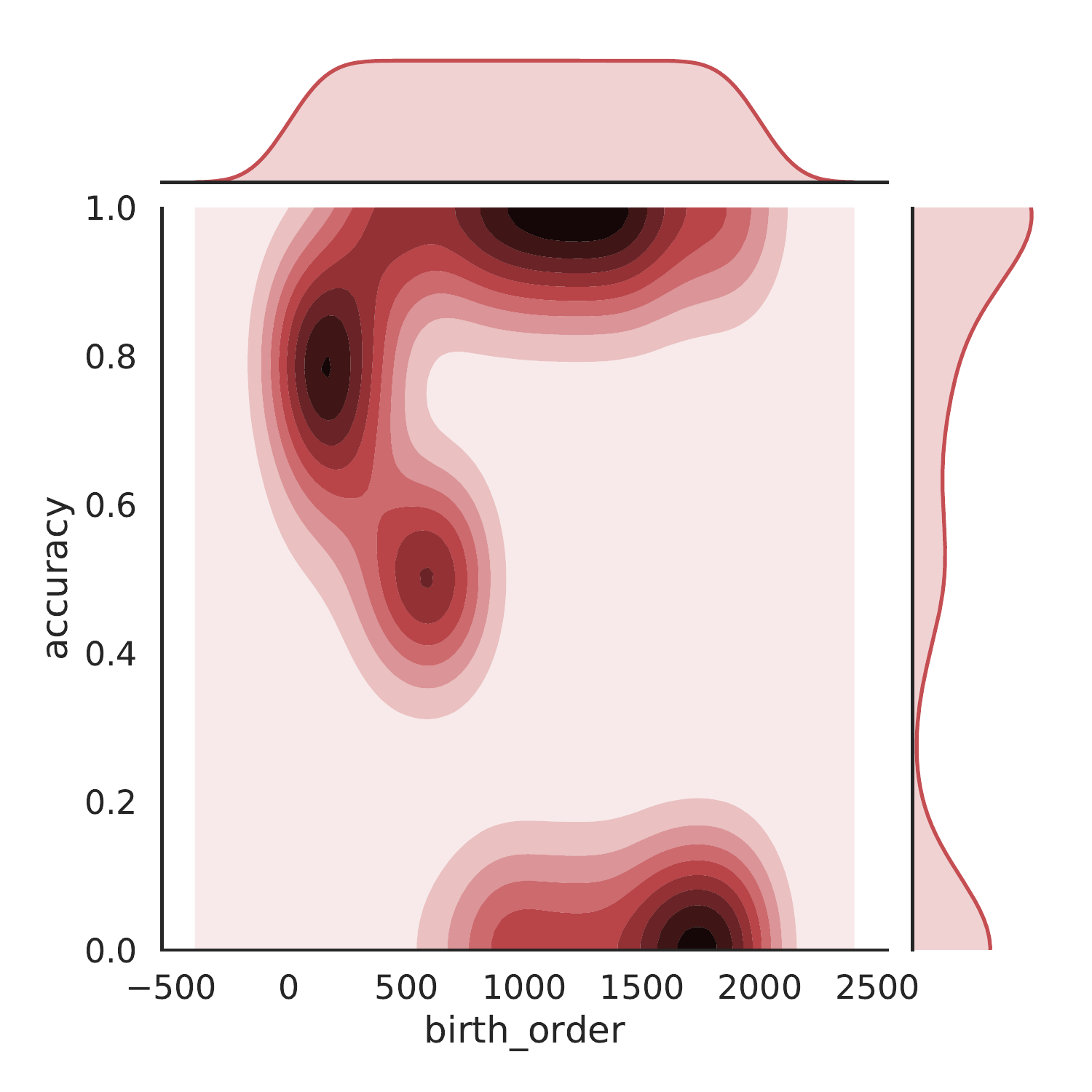}
			\label{fig:crypto_delay_accuracy_by_birth_order_pmodel}
	}
	
&

	\subfloat[Cascade-LSTM]{
		\includegraphics[width=0.24\textwidth]{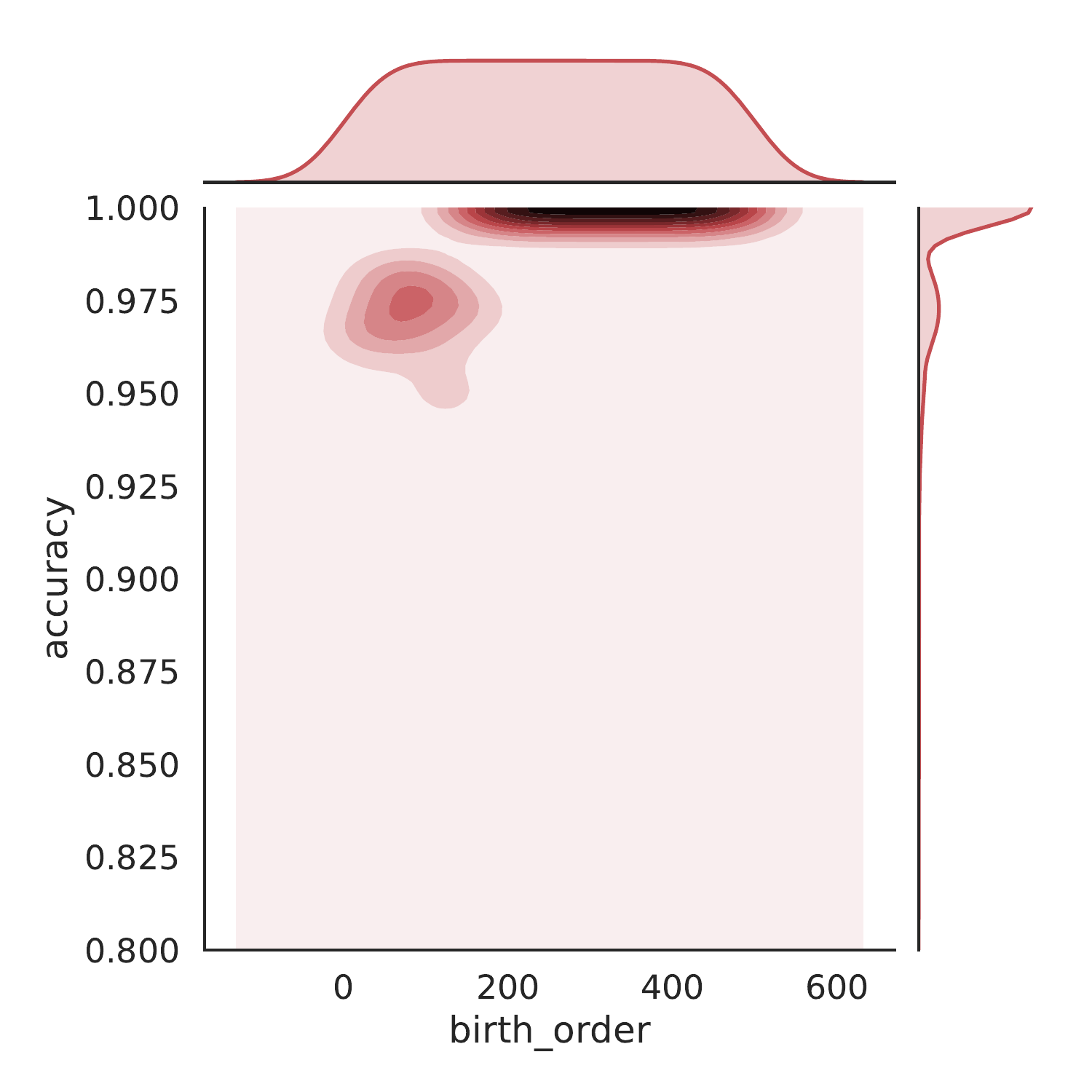}
			\label{fig:crypto_delay_accuracy_by_birth_order_clstm}
	}

\end{tabular}
	 \caption{Platform: Reddit, Accuracy Values of Speed conditioned by the level and birth order}
	 \label{fig:reddit_delay_accuracy}
	 
\end{figure}

\begin{figure}[t]
	\centering
	\centering
\begin{tabular}{cc}

	\subfloat[Baseline (level)]{
		\includegraphics[width=0.24\textwidth]{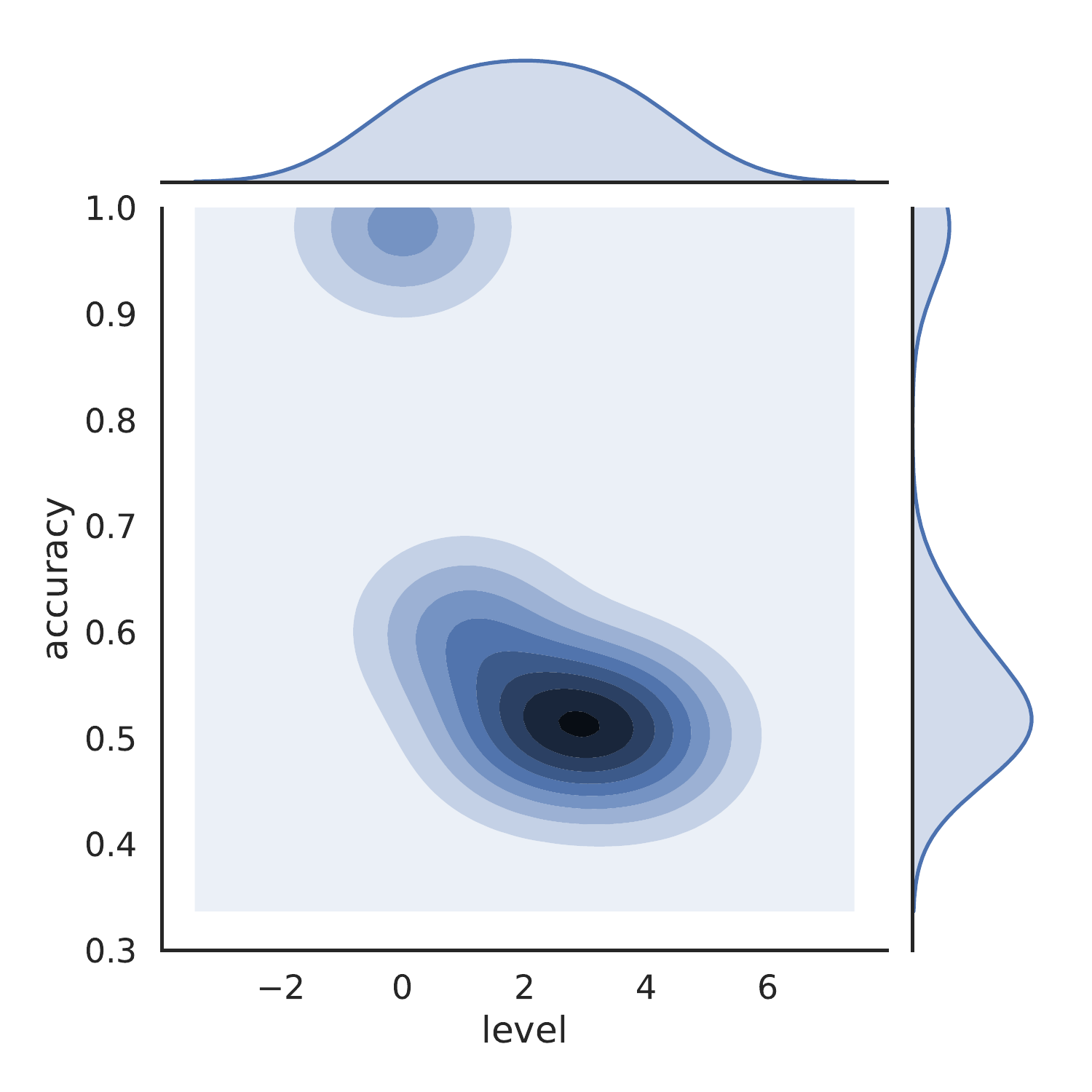}
			\label{fig:github_delay_accuracy_by_level_pmodel}
	}
	
&

	\subfloat[Cascade-LSTM]{
		\includegraphics[width=0.24\textwidth]{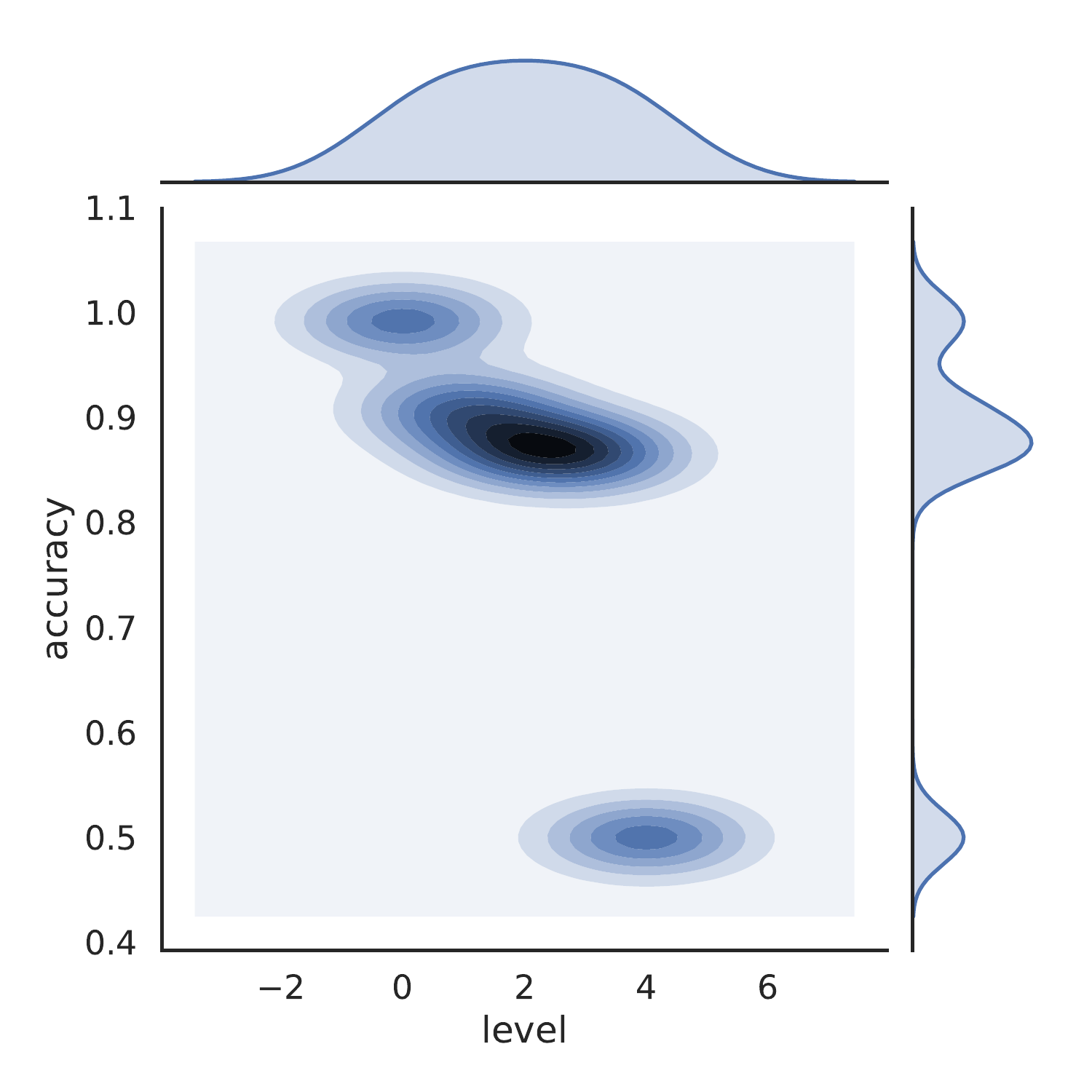}
			\label{fig:github_delay_accuracy_by_level_clstm}
	} 
	
	\\
	
		\subfloat[Baseline (birth order)]{
		\includegraphics[width=0.24\textwidth]{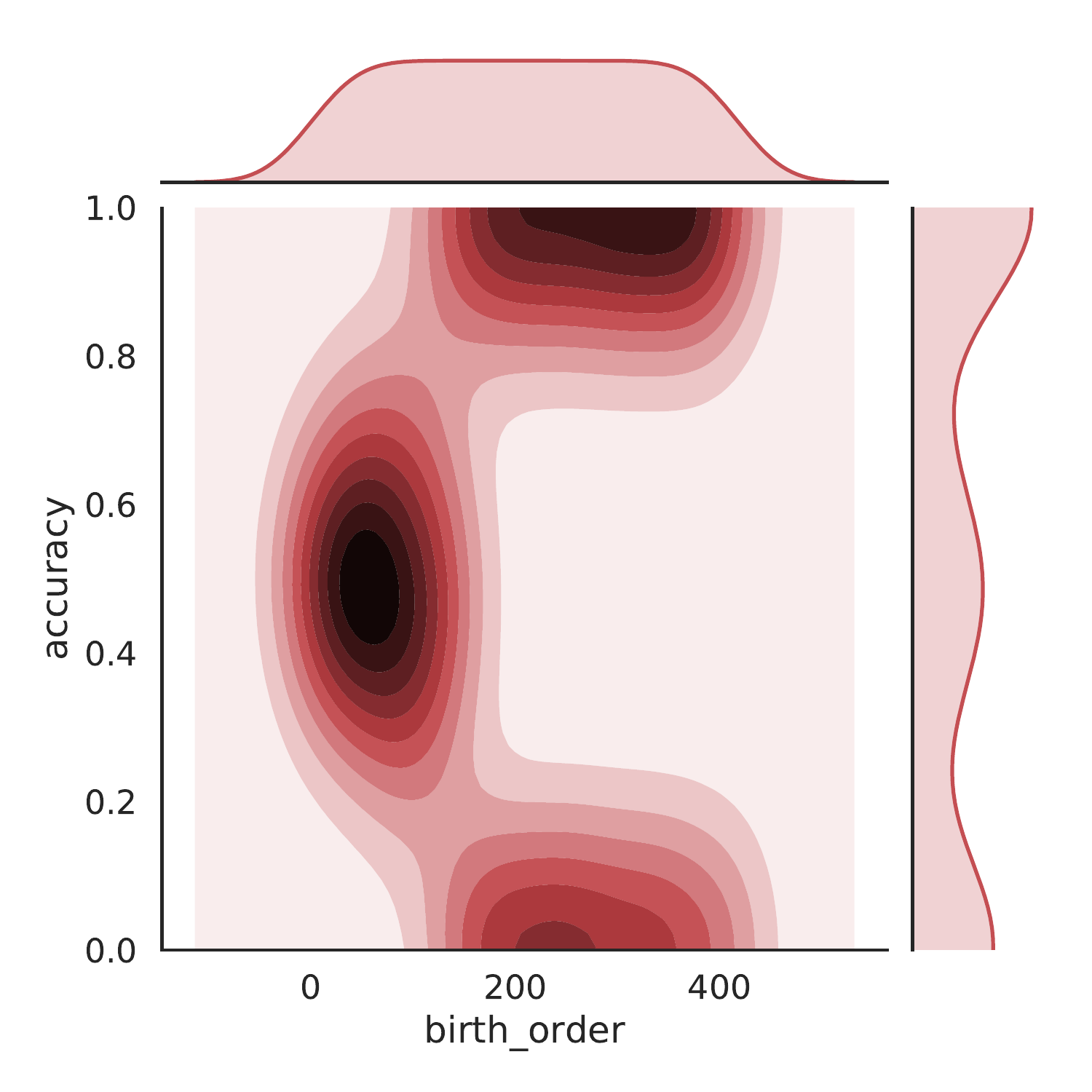}
			\label{fig:github_delay_accuracy_by_birth_order_pmodel}
	}
	
&

	\subfloat[Cascade-LSTM]{
		\includegraphics[width=0.24\textwidth]{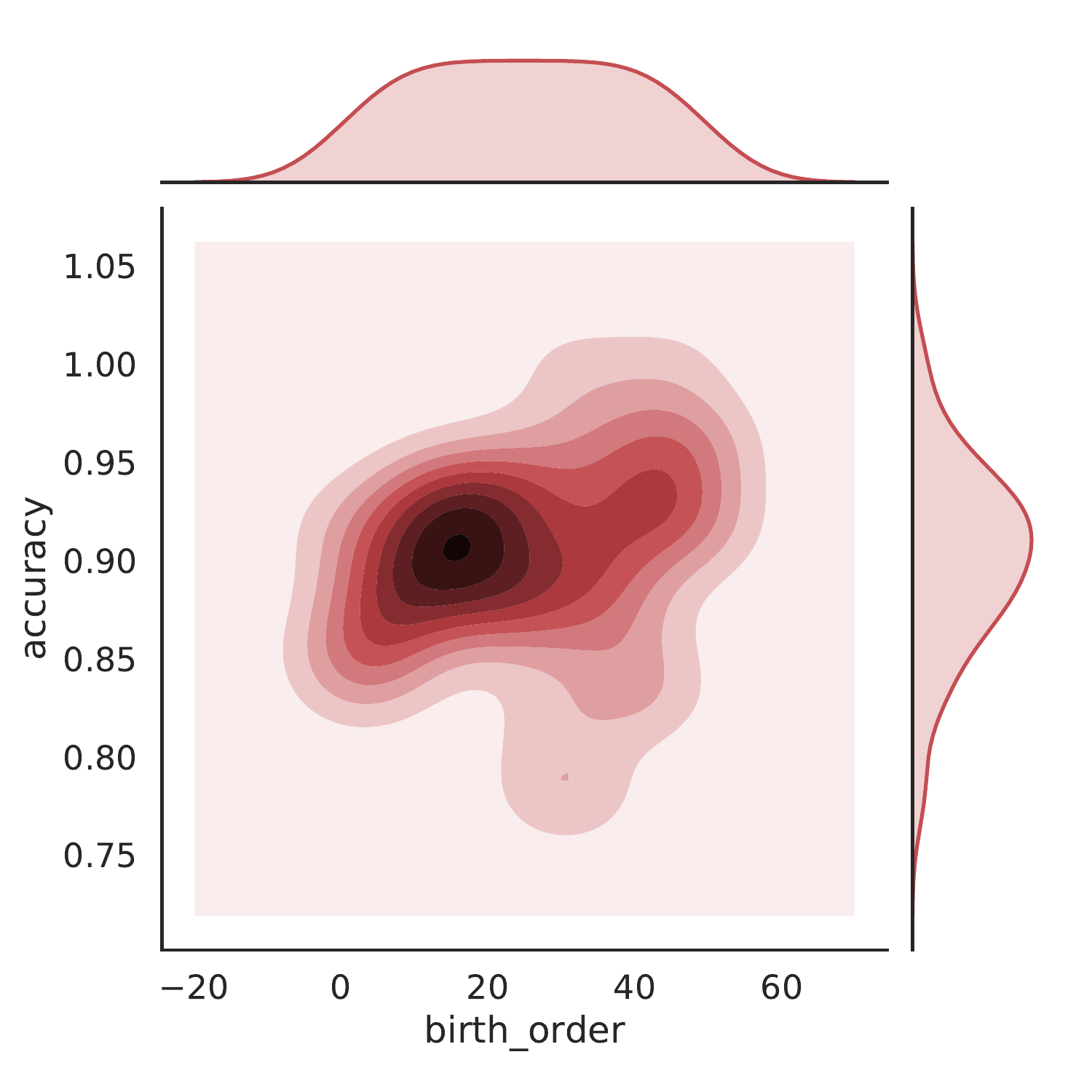}
			\label{fig:github_delay_accuracy_by_birth_order_clstm}
	}
	
\end{tabular}
	 \caption{Platform: Github, Accuracy Values of Speed conditioned by the level and birth order}
	 \label{fig:github_delay_accuracy}
	 
\end{figure}

The performance of LSTM networks is famously difficult to explain~\cite{LSTM-unreasonable}. 
In an attempt to reason about the predictability of different cascade characteristics,   
Figures~\ref{fig:degree_accuracy_reddit} and ~\ref{fig:degree_accuracy_github} present the accuracy of branch prediction as a function of level and birth order, while Figures~\ref{fig:reddit_delay_accuracy} and ~\ref{fig:github_delay_accuracy} present the accuracy of speed prediction, respectively.  
Darker colors represent higher density of accuracy values. 
The following observations can be drawn from these figures.  

As previously proven~\cite{cheng2014can}, the initial structure of the cascade is the hardest and the most important for the correct prediction of the shape of the cascade. 
As seen in Figure~\ref{fig:crypto_degree_accuracy_by_level_clstm}, our LSTM technique is particularly successful (compared to the baseline models) in accurately predicting the shape at low levels in the tree. 
Specifically, for small level values (e.g., between 0 and 25), the accuracy shown in Figure~\ref{fig:crypto_degree_accuracy_by_level_clstm} is above 60\%, while it ranges from below 40\% to 60\% for the baseline model in Figure~\ref{fig:crypto_degree_accuracy_by_level_pmodel}. 
The same behavior is evident in Github (Figures~\ref{fig:github_degree_accuracy_by_level_pmodel} and ~\ref{fig:github_degree_accuracy_by_level_clstm}) -- note the different ranges on the Y axes in this case. 

The explanation is not as clean for the impact of birth order on accuracy. As Figures~\ref{fig:crypto_delay_accuracy_by_level_pmodel} and ~\ref{fig:crypto_delay_accuracy_by_level_clstm} show, the first responses to a post are easier to predict by our LSTM approach than by the baseline approach. However, for the messages on very wide levels (i.e., when the order of the responses to a post/ comment range between 500 and 2000), our LSTM approach shows a significant drop in accuracy. LSTM recovers in its performance for the end of the spectrum: messages posted very late in the discussion -- probably because those are all leaves (conversations do not continue from the 2000th post to a particular message). 

Birth order has a better impact in Github: the LSTM technique performs significantly better than the baseline method, as shown in Figures~\ref{fig:github_delay_accuracy_by_level_pmodel} and ~\ref{fig:github_delay_accuracy_by_level_clstm} (note again the different ranges on the Y axes). The reason for this different behavior in the two platforms might be due to the radically different breadths of the cascades: while there can be thousands of responses to a post in Reddit, there are at most tens of forks of a particular repository in Github.  

The most dramatic advantage of LSTM over the baseline model is shown in the prediction of late or early adopters in both platforms (Figure~\ref{fig:reddit_delay_accuracy} and ~\ref{fig:github_delay_accuracy}). 
Note the different ranges on the Y axes. 
In Github, the baseline model predicts mostly with accuracy under 60\% for all levels, while LSTM reaches over 80\%. 
The same observation when comparing accuracy across birth order values (Figures~\ref{fig:github_delay_accuracy_by_level_pmodel} and ~\ref{fig:github_delay_accuracy_by_level_clstm}): LSTM's accuracy is around 90\% for birth orders below 60, while for the baseline accuracy ranges between 20 and 80\% (Figures~\ref{fig:github_delay_accuracy_by_birth_order_pmodel} and ~\ref{fig:github_delay_accuracy_by_birth_order_clstm}).

\subsection{Predicting Cascade Structure}
\label{sec:tree-shape}

\begin{table*}[!htbp]
    \centering
    \caption{A comparison of performance metric values between Cascade-LSTM and the baseline models for the cascades generated in Reddit and Github. We calculate JS-divergence between the distributions of the structural properties on ground-truth and generated cascade trees, where JS divergence ranges 0--1 (lower is better)
    }
    \label{tab:casacde_gen_performance}
    \begin{tabular}{|l|l|r|r|r|}
    \hline
         Platform & Structural Property & \textbf{Cascade-LSTM} & Baseline (median ranked) & Baseline (lowest ranked)  \\ \hline \hline
         \multirow{5}{*}{Reddit} & Structural virality & \textbf{0.0286} & 0.0421 & 0.0584 \\ \cline{2-5}
         & Maximum depth & \textbf{0.0767} & 0.1284 & 0.1366 \\ \cline{2-5}
         & Cascade size & \textbf{0.0415} & 0.0449 & 0.0584 \\ \cline{2-5}
         & Maximum breadth & \textbf{0.0265} & 0.0395 & 0.0448 \\ \cline{2-5}
        \hline \hline
         
         \multirow{5}{*}{Github} & Structural virality & 0.0131 & 0.0125 & \textbf{0.0104}  \\ \cline{2-5}
         & Maximum depth & \textbf{0.0010} & 0.0012 & 0.0052 \\ \cline{2-5}
         & Cascade size  & 0.0069 & 0.0075 & \textbf{0.0052} \\ \cline{2-5}
         & Maximum breadth & 0.0049 & 0.0048 & \textbf{0.0025} \\ \cline{2-5}
        \hline
    \end{tabular}
\end{table*}

Our last objective is to generate cascades that are topologically accurate.
That is, we are interested in generating the exact shape of the cascade trees. 
We are basing our generation approach on the results presented in the previous section, as described in Section~\ref{sec:generative_test}. 

The results are presented in Table~\ref{tab:casacde_gen_performance}. 
We evaluate the goodness of our generated cascades with a set of measurements representing the cascade topological structure, such as size, maximum depth, maximum breadth, and structural virality~\cite{goel2015structural}.
We use JS divergence to compare distribution-level measurements (such as the distribution of cascades over size, depth, and breadth).
The implementation of the measurements we used in this study is publicly available on Github~\cite{PNNL}.
In this case, the baseline performance is related to the selection of the cascades with the least and median accuracy in the generative test. 
Our performance results are based on 500 cascades (in each platform) generated from 500 original posts made in August 2017 in Reddit and, respectively, 500 repositories randomly picked from the Github dataset.  

The following observations can be made: First, the result of the Cascade-LSTM generator is always better than the baseline in Reddit.  
Specially, Cascade-LSTM generated cascades capture the structural diversity of the originals better than baseline (JS-divergence on comparing the distribution of structural virality is 0.02 for Cascade-LSTM, which is smaller than the baseline models)

In Github, however, we observe mixed results.
We believe this is due to the fact that the baseline models perform relatively better in the prediction tasks (Table~\ref{tbl:results}), all JS-divergence values are below 0.015 (Table~\ref{tab:casacde_gen_performance}). 
Thus, the problem becomes very hard to discriminate among a collection of realistic cascades.

\section{Summary and Discussions}
\label{sec:summary}
This paper proposed a technique for cascade generation that combines a probabilistic generative model with an LSTM-based discriminator. 
We tested our technique on two different social collaborative platforms, Reddit and Github. 
In Reddit cascades are large, fast, tall and wide. 
In Github, cascades are shallow and slow. 
Our results show that our techniques can generate accurate cascade topological structures. 

While our cascades were by definition made of posts/repositories, we note that our approach has predictive power for inferring user actions as well. 
Specifically, by predicting the temporal position of messages/repositories in the cascade tree, we can infer information about the temporal role of their authors. 
That is because our representation of the cascades as message trees allows for identifying situations when a user posts multiple messages at different times and places in the conversation thread (for Reddit) or fork tree (for Github). 
This is a behavior implicit in discussion forums such as Reddit and not common in platforms such as Twitter. 
An application of the ability of predicting temporal user roles is in developing intervention techniques for, for example, targeting users with fact-checking updates. 

We decided to represent sets of cascades as blocks of bricks instead of generating cascades sequentially (as a sequence of bricks) because of two reasons. 
First, in blocks we could capture related cascades, for example, that respond to the same exogenous events (such as legislature regulating the digital currencies). 
In this way, different discussion threads on Reddit, specific perhaps to individual coins, can be captured together in a block. Second, the block has the ability to capture users common to multiple cascades in the same data structure. 
This, we believe, is a supporting design decision for future work that includes identifying the users who will take part in a cascade. 

In this study, we only use learned models of two spatio-temporal properties (branching factor and the pace of adoption) of the cascades to perform the generative test.
In the future, we plan to develop rich artifacts to support the generative task (e.g., predict the semantic structure of the content being diffused).

\ignore{
There are two different kinds of spreaders, the one who originate the cascade (who do not adopt), and the ones who spread the content as a post-effect of an adoption.
Our objective is to differentiate them from the individuals who only adopt (non-spreaders).

In the second prediction task, we discriminate the nodes (i.e., only adopters) by the delay of adoption.
We ask how fast an individual reacts given the prior state of the cascade.
Earlier, we note that cascades are different in the pace of adoption (Figures~\ref{fig:crypto_cascade_delay} and ~\ref{fig:github-fork_cascade_delay}).
Therefore, we define two roles of an individual based on the pace of adoption relative to a cascade, namely early and late adopters.
We predict an individual is an early adopter if she does not exceed the median delay of adoption observed in the cascade, if otherwise an late adopter.
Note that, an individual could participate multiple-times in a cascade under different roles, such that we are able to discriminate the \textit{temporal role} of a node.
}

\bibliographystyle{ACM-Reference-Format}
\bibliography{acmart}

\end{document}